\documentclass[12pt,a4paper,aps,preprint,superscriptaddress,nofootinbib]{revtex4-1}
\usepackage[utf8]{inputenc}
\usepackage{graphicx}
\usepackage{amssymb}
\usepackage{textcomp}
\usepackage{amsmath}
\usepackage{tabularx}
\usepackage{bm}
\usepackage{times}
\usepackage{color}
\usepackage{slashed}
\usepackage{multirow}
\usepackage{verbatim}
\usepackage{cancel}
\usepackage{subfigure}
\usepackage{ulem}

\usepackage[colorlinks=true, pdfstartview=FitV, linkcolor=blue, citecolor=blue, urlcolor=blue]{hyperref}
\allowdisplaybreaks[4]

\def\lsim{\mathrel{\raise.3ex\hbox{$<$\kern-.75em\lower1ex\hbox{$\sim$}}}}
\def\gsim{\mathrel{\raise.3ex\hbox{$>$\kern-.75em\lower1ex\hbox{$\sim$}}}}

\newcommand{\eV}{{\rm eV}}
\newcommand{\keV}{{\rm keV}}
\newcommand{\MeV}{{\rm MeV}}
\newcommand{\GeV}{{\rm GeV}}

\newcommand{\Tr}{{\rm Tr}}

\definecolor{orange}{rgb}{1,0.5,0}

\begin{document}

\title{Revisiting the gamma-ray signal from light ALPs emitted during the evaporation of memory-burdened PBHs}

\author{Cheng-Rui Jiang}
\email{jiangcr@mail.nankai.edu.cn}
\affiliation{
School of Physics, Nankai University, Tianjin 300071, China
}
\author{Tong Li}
\email{litong@nankai.edu.cn}
\affiliation{
School of Physics, Nankai University, Tianjin 300071, China
}

\begin{abstract}
The memory-burden effect was recently proposed to take into account the information back-reaction in the evaporation of primordial black holes (PBHs).
In this work, we systematically explore the axion-like particle (ALP) production via the evaporation of memory-burdened PBHs and the subsequent ALP-photon conversion in cosmic magnetic fields. The PBH
evaporation progress is divided in the semiclassical phase of Hawking radiation and the quantum memory-burden phase. The time evolution of PBH mass and the gamma-ray flux after the conversion of PBH-emitted ALP are then evaluated by considering the memory-burden induced PBH entropy suppression. Furthermore, we investigate the detection prospects of this evaporation-conversion signal in next-generation satellite telescopes, including AMEGO, e-ASTROGAM and MAST. The detectable PBH mass range is extended down to masses much smaller than $10^{15}$ g.
Applying the Fisher forecasting method, we derive the projected sensitivity for PBH fraction $f_{\rm PBH}$ and ALP-photon coupling $g_{a\gamma\gamma}$.
\end{abstract}

\maketitle

\tableofcontents

%%%%%%%%%%%%%%%%%%%%%%%%%%%%%%%%
\section{Introduction}
\label{sec:Intro}
%%%%%%%%%%%%%%%%%%%%%%%%%%%%%%%%

Primordial black holes (PBHs) have been fascinatingly discussed for several decades, particularly in the context of dark matter (DM) and gravitational wave searches. PBHs may form via the gravitational collapse of spatially localized density perturbations in the early universe and are considered potential macroscopic DM candidates~\cite{Zeldovich:1967lct,Carr:1974nx,Carr:1975qj,Khlopov:2008qy,Barrau:2003xp} (see Ref.~\cite{Carr:2021bzv} for a recent review). Hawking's famous discovery tells that PBHs would thermally radiate elementary particles with masses below the effective temperature~\cite{Hawking:1974rv,Hawking:1975vcx}.
The Hawking radiation of PBHs is conventionally described within the semiclassical framework.
In this standard scenario, PBHs remain classical over their whole lifetime and those with masses below $M_{\rm PBH} \simeq 10^{15}$ g would have completely evaporated by now. Light PBHs cannot thus be considered the sole DM component, even though they may still have important implications for various cosmological phenomena~\cite{Carr:2020gox}.

Recently, Dvali et al. proposed the ``memory burden'' effect in PBH evaporation~\cite{Dvali:2018xpy,Dvali:2020wft,Dvali:2024hsb}. According to Dvali et al., a black hole is not merely a self-similar object that evaporates thermally as in the standard semiclassical picture. Rather, it is a system with an enormous information-storage capacity. During Hawking evaporation, the quantum information initially encoded in nearly gapless ``memory modes'' must be redistributed among fewer remaining degrees of freedom. As the black hole loses mass and entropy, this stored information back-reacts on the system, and increases the energetic cost of further decay. This phenomenon--the memory-burden effect--suppresses the late-time evaporation rate by the inverse power of its own entropy, i.e. $S^{-k}$ with $k>0$, and can stabilize the black hole against complete disappearance. In prototype quantum systems and in black-hole analogues, the effect becomes significant at the latest after the black hole has lost about half of its initial mass. Beyond this point, the evaporation is no longer self-similar: either the emission rate is drastically suppressed, or a new classical instability may set in. For PBHs, the memory-burden effect can dramatically extend their lifetimes and allows light PBHs that would otherwise have evaporated long ago to survive until today. This effect therefore reopens otherwise excluded mass windows for PBH DM~\cite{Thoss:2024hsr,Haque:2024eyh,Dvali:2024hsb}. The constraints on the parameter space of memory-burdened PBHs have been explored by analyzing high-energy gamma-ray or neutrinos~\cite{Chianese:2024rsn,Chianese:2025wrk,Tan:2025vxp,Dondarini:2025ktz,Tseng:2025fjf,Chaudhuri:2026ljw,Tan:2026ybz,Asl:2026rcm}.

The emission of beyond-the-Standard-Model (BSM) particles via Hawking radiation and its associated phenomenology have attracted considerable attention in recent literature~\cite{Calabrese:2021src,Li:2022jxo,Calabrese:2022rfa,CDEX:2022dda,Li:2022xqh,Jho:2022wxd,Agashe:2022phd,DeRomeri:2024zqs,Federico:2024fyt,Cheung:2025gdn,Li:2025lip,Zhu:2026ush,Jeesun:2026wcx}. The scenario in which light axion-like particles (ALPs, denoted as $a$)~\cite{Peccei:1977hh,Peccei:1977ur,Weinberg:1977ma,Wilczek:1977pj,Dimopoulos:1979pp,Tye:1981zy,Zhitnitsky:1980tq,Dine:1981rt,Holdom:1982ex,Kaplan:1985dv,Srednicki:1985xd,Flynn:1987rs,Kamionkowski:1992mf,Berezhiani:2000gh,Hsu:2004mf,Hook:2014cda,Alonso-Alvarez:2018irt,Hook:2019qoh} are emitted from the Hawking radiation of asteroid-mass PBHs (referred as ``PBH ALP'') and the relevant detection in terrestrial detectors were once proposed by different authors~\cite{Li:2022xqh,Jho:2022wxd,Agashe:2022phd}.
We also suggested the gamma-ray production from ``evaporation-conversion'' mechanism~\cite{Li:2025lip} via the extremely weak ALP-photon coupling
\begin{eqnarray}
-{1\over 4}g_{a\gamma\gamma} a F^{\mu\nu} \tilde{F}_{\mu\nu}\;,
\label{eq:ALP-photon}
\end{eqnarray}
where $F_{\mu\nu}$ is the SM electromagnetic field strength and $\tilde{F}_{\mu\nu}$ is its Hodge dual. After being emitted by PBHs, ALPs propagate through cosmic magnetic fields and can be converted into photons which could be probed by satellite telescopes. In these articles, the mass of PBHs was restricted to be essentially larger than about $10^{15}$ g. The emitted ALPs can reach energies peaking around a few hundred MeV from the evaporation of PBHs with asteroid-scale masses of $M_{\rm PBH}\gtrsim 10^{15}$ g. These primary ALPs are subsequently converted to secondary gamma-rays that gain sufficient kinetic energy to be detected by satellite telescopes or terrestrial detectors on Earth.

In this work, subject to the recently proposed memory-burden effect, we present a comprehensive re-evaluation of the ALP production mechanism originating from the evaporation of PBHs and the subsequent ALP-photon conversion in cosmic magnetic fields.
Our analysis builds upon the standard Hawking evaporation framework but explicitly incorporates the transition from the semiclassical phase via thermal radiation to the memory-burden phase which is governed by quantum back-reaction effects. We model this transition by introducing a memory-burden induced suppression factor into the PBH entropy evolution. This implementation directly modifies the mass-loss rate and the resulting particle spectra. The inclusion of memory-burden effect allows us to explore the PBHs with masses much lower than $10^{15}$ g. We compute the time-dependent mass evolution of PBHs across both evaporation phases and track the ALP emission spectrum.
Subsequently, we evaluate the ALP-to-photon conversion probability in large-scale cosmic magnetic fields. The resulting gamma-ray flux is derived by convolving the PBH-produced ALP spectrum with the conversion probability, properly accounting for cosmological redshift, PBH spin, and the memory-burden induced modifications to the emission profile. Importantly, the memory-burden effect extends the PBH lifetime and reshapes the energy distribution of emitted particles, thereby opening new parameter space for light PBHs that would otherwise have fully evaporated before the present epoch. This allows us to explore novel regions of the PBH mass-fraction $f_{\rm PBH}$ plane and ALP parameter space that are inaccessible under standard Hawking evaporation assumption.

To assess the observational viability of this evaporation-conversion signal, we evaluate detection prospects for next-generation satellite-borne telescopes, specifically AMEGO~\cite{AMEGO:2019gny,Kierans:2020otl}, e-ASTROGAM~\cite{e-ASTROGAM:2016bph,e-ASTROGAM:2017pxr} and MAST~\cite{Dzhatdoev:2019kay} experiments. These instruments, sensitive to the MeV-GeV range where the ALP-converted gamma-ray signal is expected to peak, offer unprecedented capabilities for probing PBH-induced emissions.
We employ the Fisher forecasting method to quantify the constraining power of these future missions. Finally, we derive the projected sensitivity for PBH fraction $f_{\rm PBH}$ and ALP-photon coupling $g_{a\gamma\gamma}$.

This paper is organized as follows. In Sec.~\ref{sec:PBH}, we overview the mechanism of memory-burdened PBH and the time evolution of PBH mass. The flux of emitted ALP from memory-burdened PBH evaporation will be shown in galactic and extragalactic regions. The conversion of PBH-emitted ALP under magnetic field and the resultant gamma-ray flux
are then calculated in Sec.~\ref{sec:Conversion}. We show the prospective detection results of memory-burdened PBH and ALP-photon coupling in Sec.~\ref{sec:Sensitivity}. Our conclusions are drawn in Sec.~\ref{sec:Con}.

%%%%%%%%%%%%%%%%%%%%%%%%%%%%%%%%%%%
\section{Light axion-like particle from memory-burdened PBH}
\label{sec:PBH}
%%%%%%%%%%%%%%%%%%%%%%%%%%%%%%%%%%%

The evaporation of PBHs is conventionally considered to occur exclusively via Hawking radiation within the semiclassical framework. Any particle species with a mass below the following threshold temperature $T_{\rm PBH}$ can in principle be emitted through PBH evaporation~\cite{Hawking:1975vcx,Page:1976df,Page:1977um,MacGibbon:1990zk,MacGibbon:1991tj}
\begin{eqnarray}
k_BT_{\rm PBH}=\frac{\hbar c ^3}{8\pi GM_{\rm PBH}}\;,
\end{eqnarray}
where $k_B$ is the Boltzman constant, $G$ is the Newtonian constant of gravitation, and $M_{\rm PBH}$ denotes the PBH mass.
The PBH emission rate of particle $i$ can be expressed as~\cite{Hawking:1971ei,Page:1976df,Page:1976ki}
\begin{eqnarray}
\frac{d^2N_i}{dEdt}=\frac{g_i}{2\pi}\frac{\Gamma_i(E,M_{\rm PBH},a^\ast)}{e^{E/T_{\rm PBH}}-(-1)^{2s_i}}\;,
\end{eqnarray}
where $E$ is the emitted particle energy, $\Gamma_i$ is the so-called ``graybody'' factor describing the probability of elementary particle $i$ escaping the PBH gravitional well, $a^\ast$ is the PBH spin, $g_i$ and $s_i$ denote the degree of freedom and spin of emitted particle $i$, respectively.
The mass loss rate of an evaporating PBH is then given by~\cite{Page:1976df,MacGibbon:1990zk,MacGibbon:1991tj,Haque:2024eyh}
\begin{eqnarray}
\frac{dM_{\rm PBH}(t)}{dt}
=-\sum_i \int_0^\infty \frac{d^2N_i}{dEdt} EdE
\simeq -\frac{\mathcal{G}g_\ast(T_{\rm PBH})}{30720\pi G^2M^2_{\rm PBH}(t)}\;,
\label{eq:massloss}
\end{eqnarray}
where $\mathcal{G}$ denotes the graybody factor, and $g_\ast(T_{\rm PBH})$ is the number of degrees of freedom at the PBH temperature $T_{\rm PBH}$.
After integrating Eq.~(\ref{eq:massloss}) from the PBH formation time $t_{0}$, we get the PBH mass at time $t$ as
\begin{eqnarray}
M_{\rm PBH}(t)=M_{\rm PBH}[1-\Gamma^0_{\rm PBH}(t-t_0)]^{1/3}\;,
\label{eq:Mt}
\end{eqnarray}
where $M_{\rm PBH}$ denotes the initial mass of PBH, and the decay width associated with PBH evaporation is $\Gamma^0_{\rm PBH}=\mathcal{G} g_\ast/(10240 \pi G^2 M_{\rm PBH}^3)$. Suppose the PBH lifetime is much longer than $t_0$, the lifetime becomes
\begin{eqnarray}
\tau_{\rm PBH}\simeq{\Gamma_{\rm PBH}^0}^{-1}=\frac{10240\pi G^2M_{\rm PBH}^3}{\mathcal{G}g_\ast}
\approx
4.4\times 10^{17}~{\rm s}~\Big(\frac{M_{\rm PBH}}{1.24\times 10^{15}~{\rm g}}\Big)^3\;,
\end{eqnarray}
where we take the graybody factor $\mathcal{G}=27/4$~\cite{Baldes:2020nuv,Cheek:2021odj} and the number of degrees of freedom in the SM at a high temperature $g_\ast\simeq 106.75$~\cite{Baldes:2020nuv,Haque:2024eyh}. It turns out that the PBHs with an initial mass less than about $10^{15}~{\rm g}$ would have completely evaporated by today. In the realistic calculation of this work, we employ the public code \texttt{BlackHawk\_v2.3}~\cite{Arbey:2019mbc,Arbey:2021mbl} to generate the above number density of emitted particle and the running PBH mass by taking into account new degree of freedom. The spin of PBH is taken as $a^\ast=0$ or $a^\ast=1$ in \texttt{BlackHawk\_v2.3}. We also assume a monochromatic
mass distribution for PBHs.

\begin{figure}[htbp]
\centering
\includegraphics[width=0.7\linewidth]{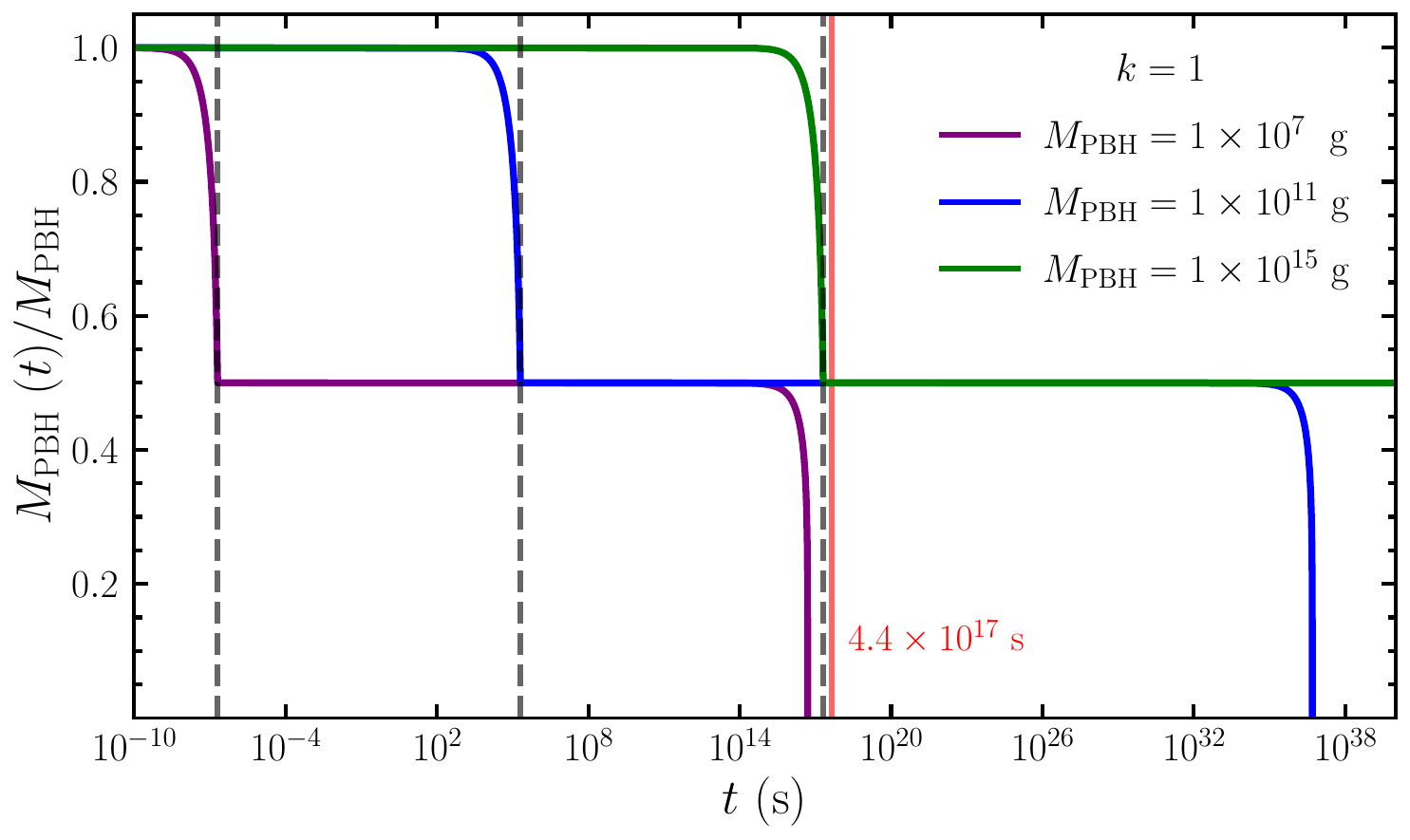}
\caption{The normalized time evolution of PBH mass $M_{\rm PBH}(t)/M_{\rm PBH}$ with $q=0.5$ and $k=1$ for $M_{\rm PBH}=10^{7}~{\rm g}$ (purple line), $10^{11}~{\rm g}$ (blue line) and $10^{15}~{\rm g}$ (green line). The red vertical line denotes the current age of the universe.
}
\label{fig:M_t}
\end{figure}

The memory-burdened PBH is characterized by taking into account the back-reaction effect when the PBH mass reaches a certain fraction
of its initial value during evaporation. This memory-burden effect suggests the
quantum modes with the entropic degrees of freedom of a black hole. As a result, the semiclassical Hawking radiation is not valid any more after this time scale and the PBH emission rate is suppressed by its entropy to the power of a positive integer. The PBH evaporation progress is then divided into two phases. The first one is the phase of semiclassical Hawking radiation following the above characteristics. Then, the PBH enters the memory-burden phase in which the quantum memory effect decelerates the evaporation process. From Eq.~(\ref{eq:Mt}), one can obtain the duration of the semiclassical phase as
\begin{eqnarray}
t_q=\tau_{\rm PBH}(1-q^3)\;,
\end{eqnarray}
where $q$ denotes the fraction of the initial PBH mass at which the semiclassical regime comes to the end.
The PBH mass at the begin of memory-burden phase is then given as
\begin{eqnarray}
M_{\rm PBH}^{\rm mb}=q M_{\rm PBH}\;,
\end{eqnarray}
where we take $q=0.5$ as a representative value for which the memory-burden effect sets in. Once $t$ exceeds $t_q$, the information accumulated on the PBH horizon gives rise to a back-reaction and reduces the mass-loss rate
\begin{eqnarray}
\frac{dM_{\rm PBH}^{\rm mb}}{dt}=\frac{1}{S^k(M_{\rm PBH})}\frac{dM_{\rm PBH}}{dt}\;,
\label{eq:mmbmassloss}
\end{eqnarray}
where $k$ is a positive integer and $S$ is the PBH entropy defined as
\begin{eqnarray}
S(M_{\rm PBH})=4\pi GM^2_{\rm PBH}\simeq2.6\times10^{40}\Big(\frac{M_{\rm PBH}}{10^{15}~{\rm g}}\Big)^2\;.
\end{eqnarray}
After integrating Eq.~(\ref{eq:mmbmassloss}) from $t_q$, we get the time evolution of the PBH mass during the memory-burden phase
\begin{eqnarray}
M^{\rm mb}_{\rm PBH}(t)=M_{\rm PBH}^{\rm mb}[1-\Gamma^{(k)}_{\rm PBH}(t-t_q)]^{1/(3+2k)}\;,
\label{eq:mmb_t}
\end{eqnarray}
where
\begin{equation}
\Gamma^{(k)}_{\rm PBH}=\frac{\mathcal{G}g_\ast}{7680\pi}2^k (3+2k)M_{\rm p}\Big(\frac{M_{\rm p}}{M^{\rm mb}_{\rm PBH}}\Big)^{3+2k},
\end{equation}
with $M_{\rm p}=(8\pi G)^{-1/2}$ being the reduced Plank mass. We fix $k=1$ in the following analysis. In Fig.~\ref{fig:M_t} as illustration, we show the normalized time evolution of PBH mass with $q=0.5$ and $k=1$ for $M_{\rm PBH}=10^{7}~{\rm g}$ (purple line), $10^{11}~{\rm g}$ (blue line) and $10^{15}~{\rm g}$ (green line). Since the evaporation time scales as $1/M_{\rm PBH}^3$, one can see that the PBHs with a smaller initial mass enter the memory-burden phrase at earlier times. After entering the memory-burden phase, the PBH mass stays approximately constant over the entire age of the universe for $M_{\rm PBH}$ as low as about $10^{7}~{\rm g}$. The PBHs with mass larger than about $10^{15}~{\rm g}$ would not enter the memory-burden phase.

\begin{figure}
\centering
\includegraphics[scale=1,width=0.9\linewidth]{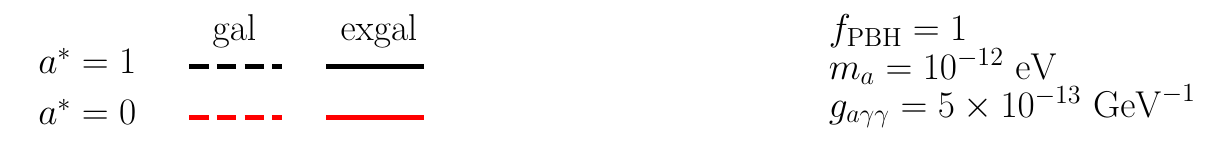}
\includegraphics[scale=1,width=0.49\linewidth]{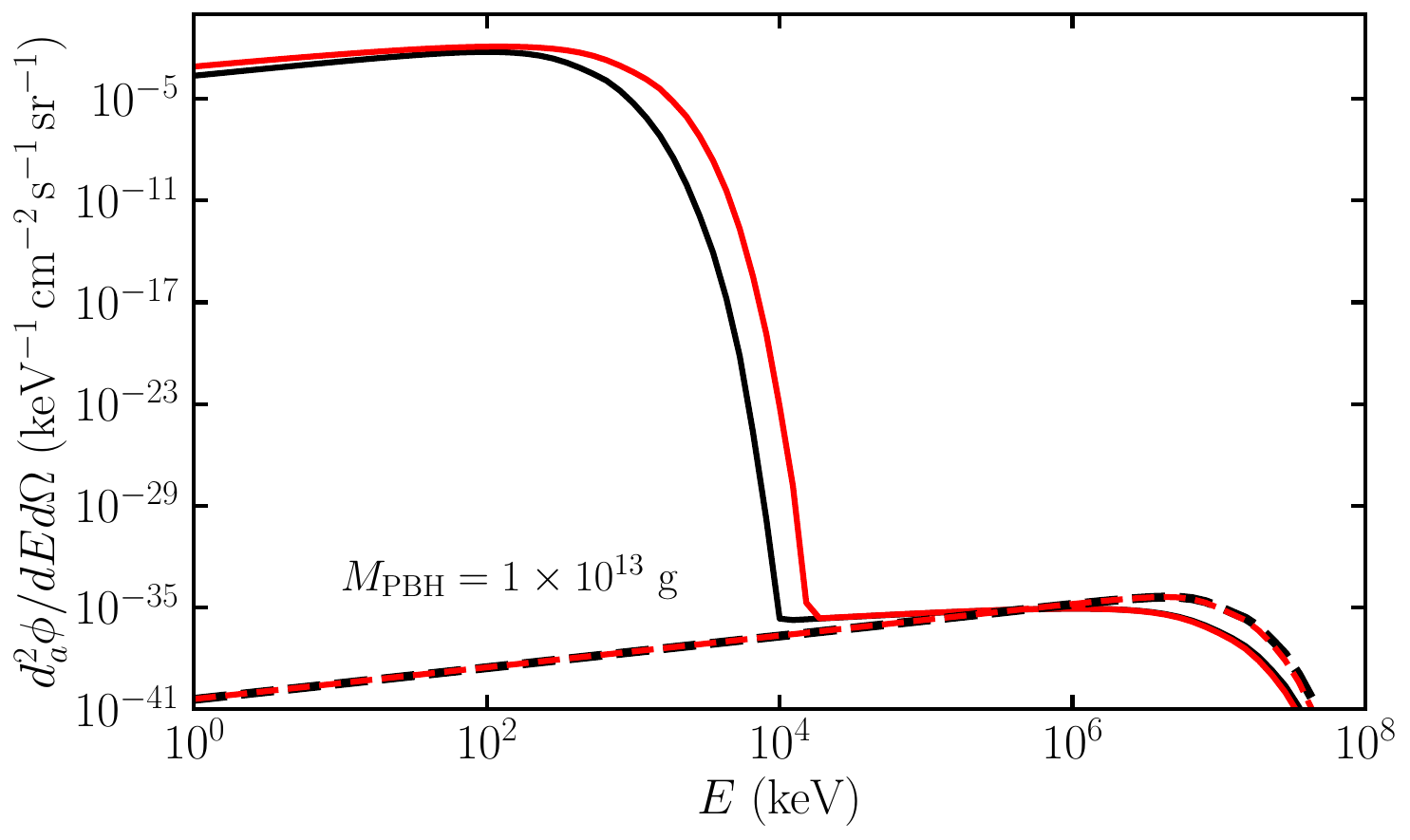}
\includegraphics[scale=1,width=0.49\linewidth]{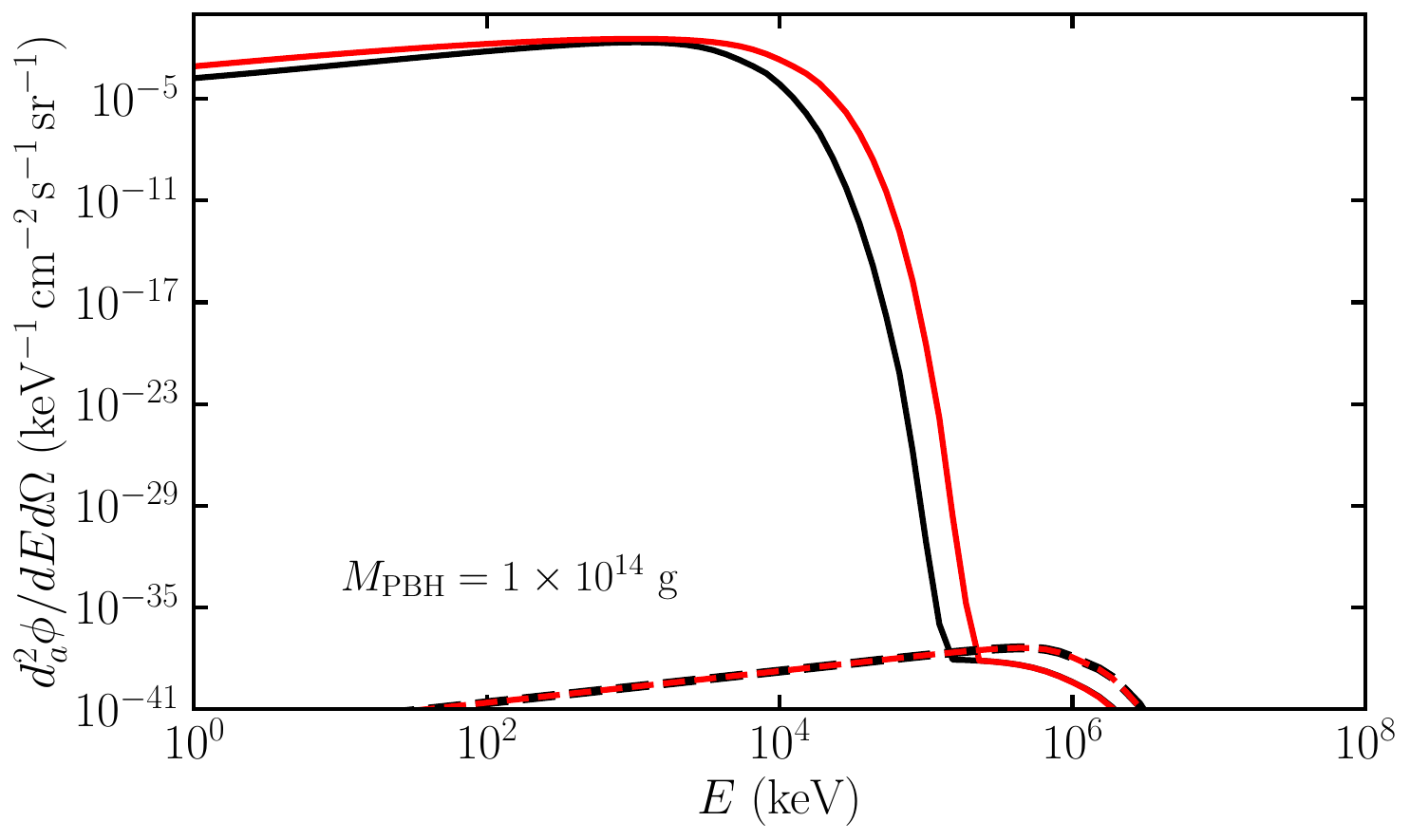}
\includegraphics[scale=1,width=0.49\linewidth]{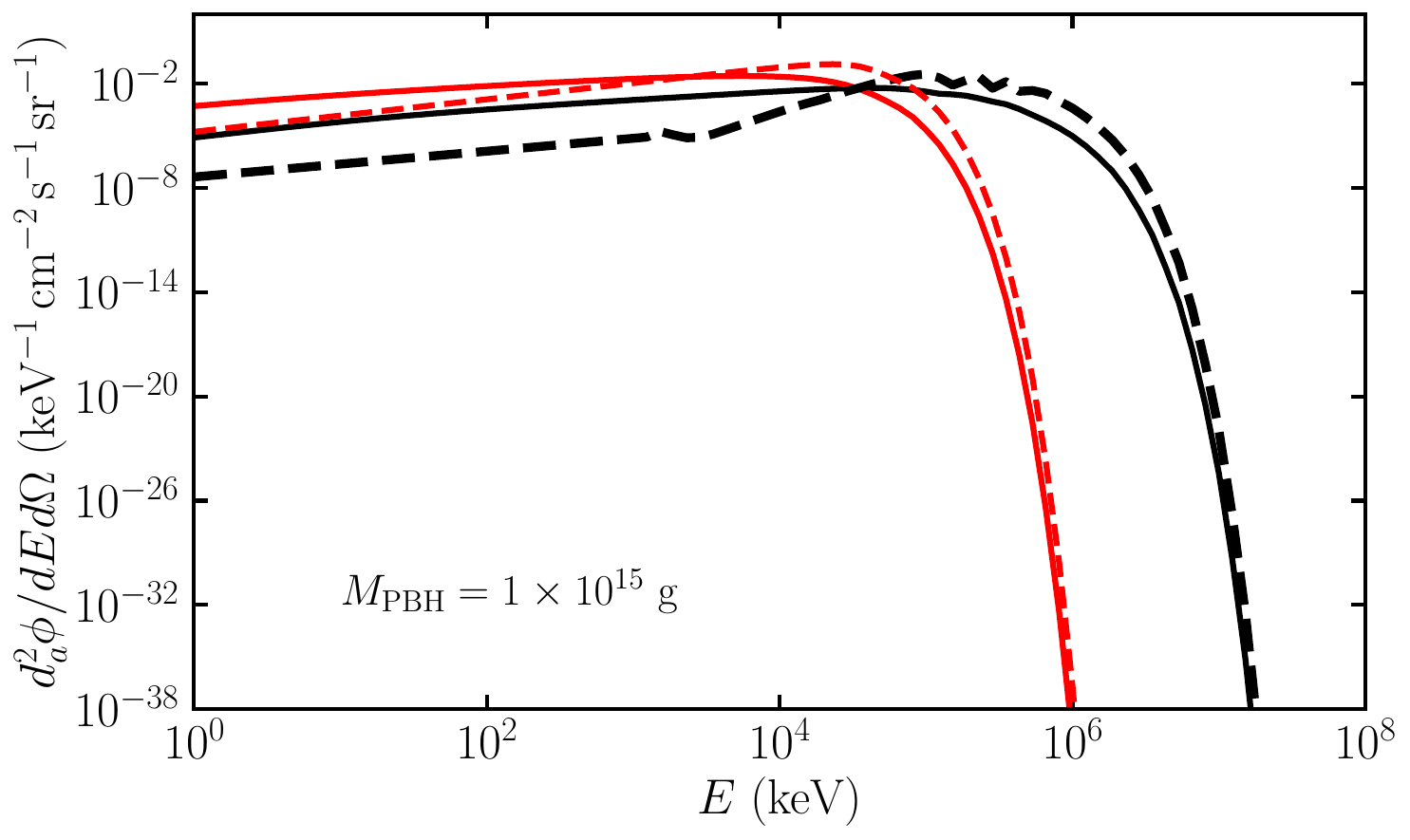}
\includegraphics[scale=1,width=0.49\linewidth]{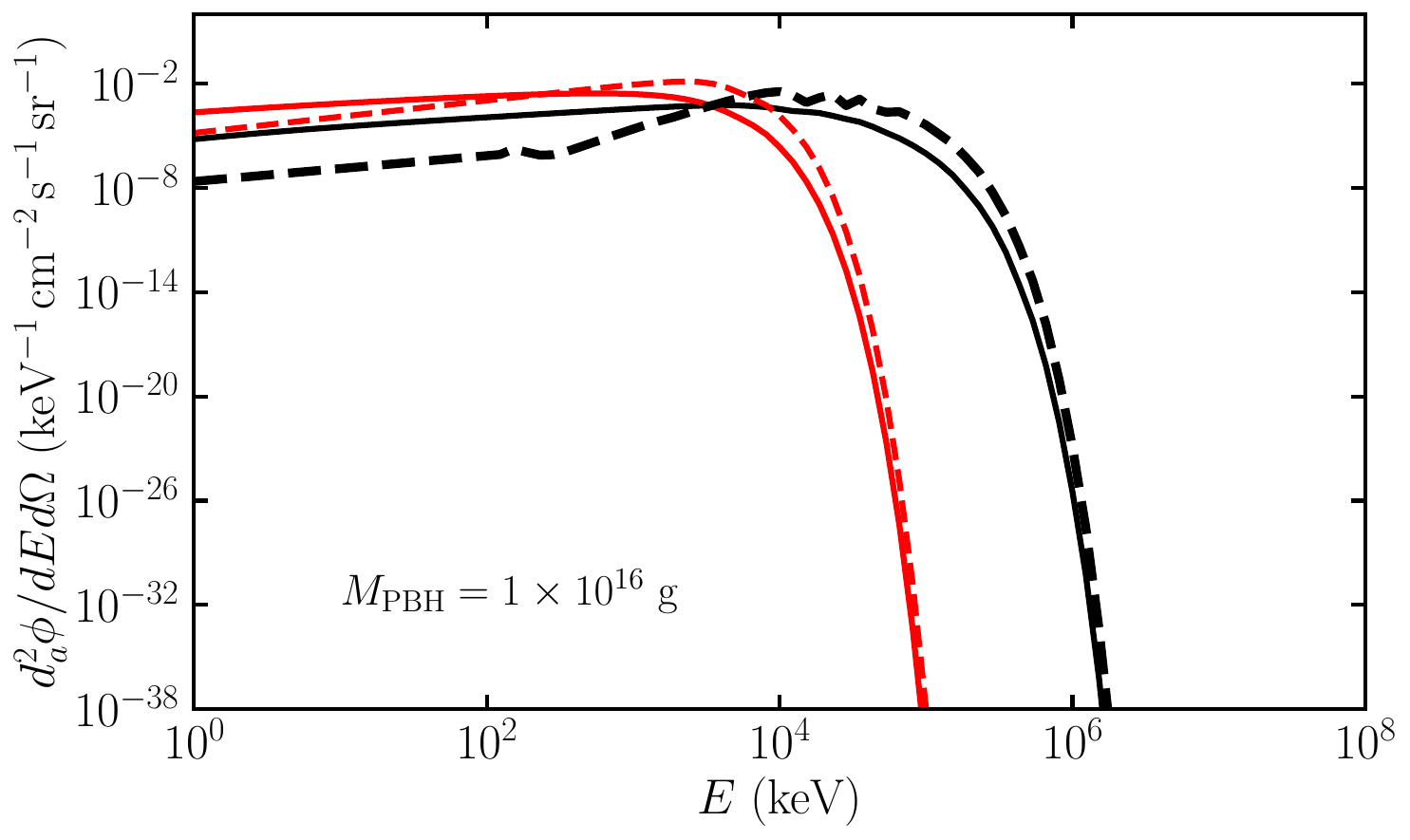}
\caption{The flux of ALPs from memory-burdened PBH evaporation for $M_{\rm PBH}=1\times 10^{13}$ g (top left panel), $M_{\rm PBH}=1\times 10^{14}$ g (top right panel), $M_{\rm PBH}=1\times 10^{15}$ g (bottom left panel) and $M_{\rm PBH}=1\times 10^{16}$ g (bottom right panel), assuming $f_{\rm PBH}=1$, $m_a = 10^{-12}~{\rm eV}$ and $g_{a\gamma\gamma}=5\times 10^{-13}~{\rm GeV}^{-1}$. Both galactic (dashed lines) and extragalactic (solid
lines) PBH contributions are included with $a^\ast=0$ (red) or $a^\ast= 1$ (black).
}
\label{fig:PBH_Spectrum}
\end{figure}

The PBH emission rate becomes
\begin{eqnarray}
\frac{d^2N}{dEdt} =\frac{d^2N}{dEdt}(M_{\rm PBH}(t))
\end{eqnarray}
in the semiclassical phase $t\le t_q$, and
\begin{eqnarray}
\frac{d^2N^{\rm mb}}{dEdt}=\frac{1}{S^k(M^{\rm mb}_{\rm PBH}(t))}\dfrac{d^2N}{dEdt}(M_{\rm PBH}^{\rm mb}(t))
\end{eqnarray}
in the memory-burden phase $t>t_q$. Note that we next consider the PBH mass evolution in the calculation of emission rate.
The particle flux emitted from PBH can be divided into the
contributions of the PBHs in galactic halo and the extragalactic PBHs
\begin{equation}
\frac{d^2\Phi}{dEd\Omega}=\frac{d^2\Phi_{\rm{gal}}}{dEd\Omega}+\frac{d^2\Phi_{\rm{exgal}}}{dEd\Omega}\;,
\end{equation}
where the first term and the second term on the right-hand side correspond to the flux averaged over the whole solid angle from galactic (gal) PBHs and extragalactic (exgal) PBHs, respectively.
The evaporation spectrum from PBHs within the Milky Way is given by the PBH emission rate in the memory-burden phase as follows
\begin{eqnarray}
\frac{d^2\Phi_{\rm gal}}{dEd\Omega}=\frac{f_{\rm PBH}}{4\pi M_{\rm PBH}^{\rm mb}}\frac{d^2N^{\rm mb}}{dEdt}\int_{\rm RoI}\frac{d\Omega_s}{\Delta\Omega}\int_0^{s_{\rm max}}ds\rho[r(s,l,b)]\;,
\end{eqnarray}
where $f_{\rm PBH}$ denotes the fraction of DM composed of PBHs, $M_{\rm PBH}^{\rm mb}$ is the present PBH mass in memory-burden phase, and $\Delta\Omega$ is the integral to solid angle within the Region of Interest (RoI) given by the line-of-sight distance $s$ and the galactic coordinates $(l, b)$. The differential solid angle element turns out to be $d\Omega_s=dl db \cos(b)$. For the gamma-ray experiments that we will analyze, the emission signals are confined to the range $|l|\leq 5^\circ$ and $|b|\leq 5^\circ$~\cite{Ray:2021mxu}.
For the local DM profile, we implement the Navarro-Frenk-White (NFW) density form~\cite{Navarro:1996gj}
\begin{eqnarray}
\rho(r)=\rho_\odot\left(\frac{r}{r_\odot}\right)^{-\gamma}\left(\frac{1+r_\odot/r_s}{1+r/r_s}\right)^{3-\gamma}~~{\rm with}~\gamma=1\;,
\end{eqnarray}
where $\rho_\odot=0.4~\GeV/{\rm cm}^3$ denotes the local DM density~\cite{Salucci:2010qr}, $r_\odot=8.5$ kpc is the sun-galactic center distance~\cite{Yuan:2017ozr}, and $r_s=20$ kpc is the radius of the galactic diffusion disk. The galactocentric distance is given by $r=[r_\odot^2+s^2-2sr_\odot\cos (l)\cos (b)]^{1/2}$. The maximum line-of-sight distance $s_{\rm max}$ in the integral of this DM density distribution becomes
\begin{eqnarray}
s_{\rm max}=r_\odot\cos (b)\cos (l)+\sqrt{r_{\rm max}^2-r_\odot^2[1-\cos^2(b)\cos^2(l)]}\;,
\end{eqnarray}
with the DM halo radius $r_{\rm max}=200$ kpc~\cite{Wang:2020uvi}.
The differential flux per unit solid angle that we calculate here is essentially an angular average of a line-of-sight integral, rather
than a volume integration. The integration variable $s$ denotes the path length along a fixed direction of $(l, b)$. The integral physically represents the accumulation of the PBH-emitted particle flux along that direction.
For extragalactic PBHs at cosmological distance, we must account for redshift effects and modify the particle emission spectrum as
\begin{eqnarray}
\frac{d^2\Phi_{\rm exgal}}{dEd\Omega}=\frac{f_{\rm PBH}\rho_{\rm DM}}{4\pi M_{\rm PBH}}\int_{z=z(t_q)}^\infty\frac{dz}{H(z)}\frac{d^2N}{dEdt}\Bigg\vert_{E\to E_z}+\frac{f_{\rm PBH}\rho_{\rm DM}}{4\pi M_{\rm PBH}^{\rm mb }}\int_{z=0}^{z=z(t_q)}\frac{dz}{H(z)}\frac{d^2N^{\rm mb }}{dEdt}\Bigg\vert_{E\to E_z}\;,
\label{eq:exgal}
\end{eqnarray}
where the average DM density of the universe at the present epoch is $\rho_{\rm DM}=2.35\times 10^{-30}~{\rm g}/{\rm cm}^3$~\cite{Planck:2018vyg},
$E_z=\sqrt{(E^2-m^2)(1+z(t))^2+m^2}$ denotes the energy at the source and is related to the energy $E$ in the observer's frame by the redshift $z(t)$ for a particle with mass $m$, $H(z)=H_0\sqrt{\Omega_\Lambda+\Omega_m(1+z)^3+\Omega_r(1+z)^4}$ is the Hubble expansion rate at the redshift $z$ where $\Omega_\Lambda, \Omega_m$ and $\Omega_r$ are current dark-energy, matter and radiation densities of the universe, respectively.
Note that here we split the integral over redshift $z$ into the integral in the semiclassical phase (the first term on the right-hand side of Eq.~(\ref{eq:exgal})) and the integral in the memory-burden phase (the second term).
Fig.~\ref{fig:PBH_Spectrum} shows the flux of ALPs from memory-burdened PBH evaporation for $M_{\rm PBH}=1\times 10^{13}$ g (top left panel), $M_{\rm PBH}=1\times 10^{14}$ g (top right panel), $M_{\rm PBH}=1\times 10^{15}$ g (bottom left panel) and $M_{\rm PBH}=1\times 10^{16}$ g (bottom right panel),
assuming $f_{\rm PBH}=1$, $m_a = 10^{-12}~{\rm eV}$ and $g_{a\gamma\gamma}=5\times 10^{-13}~{\rm GeV}^{-1}$. Both galactic (dashed lines) and extragalactic (solid
lines) PBH contributions are displayed with spin $a^\ast=0$ (red) or $a^\ast= 1$ (black) for comparison. For $M_{\rm PBH}\lesssim 10^{15}$ g, one can see that both the galactic and extragalactic contributions from memory-burdened PBHs are highly suppressed by the PBH entropy, much smaller than the extragalactic part in semiclassical phase. PBHs with smaller initial masses enter the memory-burden phase at an earlier time, yielding lower-energy ALP emission from a sizable extragalactic contribution. Furthermore, a larger PBH spin $a^\ast$ boosts the emission rate and advances the onset of memory-burden phase.
On the other hand, for $M_{\rm PBH}\gtrsim 10^{15}$ g, PBHs no longer enter memory-burden phase and the ALP flux doesn't suffer from entropy suppression. The ALP energy decreases as PBH mass increases. A larger PBH spin parameter $a^\ast$ enhances the emission energy.

%%%%%%%%%%%%%%%%%%%%%%%%%
\section{Conversion of PBH-emitted ALP under cosmic magnetic field and gamma-ray flux}
\label{sec:Conversion}
%%%%%%%%%%%%%%%%%%%%%%%%%

The ALPs emitted from PBHs can be efficiently converted into photons via astrophysical magnetic fields while propagating through the galaxies~\cite{Raffelt:1987im}. Once reaching the solar system, these gamma-rays may be detected by space-based telescope experiments. In this article and our earlier paper~\cite{Li:2025lip}, this gamma-ray production mechanism is referred to as the ``evaporation-conversion'' process. In this section, we will discuss this process in details and show the corresponding gamma-ray signal spectra modified by the memory-burden mechanism.

After being emitted from PBHs and during propagation, the ALPs interact with the ambient magnetic fields through the ALP-photon interaction term $-{1\over 4}g_{a\gamma\gamma} aF^{\mu\nu}\tilde{F}_{\mu\nu}$.
This coupling induces coherent oscillations between ALPs and photons under the resonant condition. As a result, the ALPs are converted into detectable photons.
For a monochromatic photon or ALP beam travelling along $x_3$ axis~\footnote{For the ALP-photon conversion issue, the $x_3$ axis is aligned with the photon propagation direction. The $x_1,x_2$ axes form the transverse plane perpendicular to the propagation direction.} in a cold plasma, their propagation can be described by the following Schr\"{o}dinger-like equation~\cite{Raffelt:1987im}
\begin{eqnarray}
    \left(i\frac{d}{dx_3}+E+\mathcal{M}\right)\left(\begin{matrix}
        A_1(x_3)\\A_2(x_3)\\a(x_3)
    \end{matrix}\right)=0\;,
\end{eqnarray}
where $A_1(x_3)$ and $A_2(x_3)$ are the photon amplitudes with polarization along the $x_1$ and $x_3$ axis, respectively, and $a(x_3)$ denotes the ALP field. $\mathcal{M}$ is the ``ALP-photon mixing matrix'' containing the ALP-photon interaction.
Under an assumption of an external homogeneous magnetic field, the previous works~\cite{Raffelt:1987im,Mirizzi:2006zy,Galanti:2022ijh} provide a commonly used ALP-photon conversion probability. However, the real astrophysical magnetic fields are far from homogeneous and instead possess highly complex structures. In the Milky Way, the magnetic field exhibits a complex multi-scale structure~\cite{Pshirkov:2011um,Jansson:2012pc}. While in galaxy clusters, observational results clearly indicate that the magnetic field exhibits turbulent behavior~\cite{1993ApJ416554T,Ensslin:2003ez,Vogt:2003su}.
To address these complex situations, we adopt a rigorous propagation matrix method and separately calculate the conversion probabilities when the PBH ALPs traverse galaxy clusters and the Milky Way.

In galaxy clusters, the magnetic field is inhomogeneous and can be described as a set of isotropic regions with randomly distributed orientations. This property indicates the presence of a significant turbulent structure in which the direction of the magnetic field varies sharply from one location to another.
Based on Ref.~\cite{Horns:2012kw}, the turbulent magnetic field in a galaxy cluster can be modeled as a set of coherent domains. Within each domain, the magnetic field strength is approximately constant and the direction is fixed. However, when crossing to an adjacent domain, the direction undergoes a sudden change and follows a random distribution.
Since the magnetic field direction is completely random, the mixing matrix $\mathcal{M}$ can be represented by a general three-dimensional matrix
\begin{eqnarray}
\mathcal{M}_k=\left(\begin{matrix}
        \Delta_{xx}&\Delta_{xy}&\Delta_{a\gamma}\sin\psi_k\\
        \Delta_{yx}&\Delta_{yy}&\Delta_{a\gamma}\cos\psi_k\\
        \Delta_{a\gamma}\sin\psi_k&\Delta_{a\gamma}\cos\psi_k&\Delta_a
\end{matrix}\right)\;,
\end{eqnarray}
where the subscript $k$ distinguishes different coherent domains, and $\psi_k$ represents the random magnetic field orientation within each domain.
The matrix elements are given by
$\Delta_{xx}=\Delta_\parallel\sin^2\psi_k+\Delta_\bot\cos^2\psi_k$, $\Delta_{yy}=\Delta_\parallel\cos^2\psi_k+\Delta_\bot\sin^2\psi_k$, $\Delta_{xy}=\Delta_{yx}=(\Delta_\parallel-\Delta_\bot)\sin\psi_k\cos\psi_k$, where $\Delta_\parallel=\Delta_{\rm pl}+7/2\Delta_{\rm QED}$ and $\Delta_\bot=\Delta_{\rm pl}+2\Delta_{\rm QED}$ are two wavenumbers.
For the considered energy window of $\mathcal{O}(1-100)~{\rm MeV}$ in our case, the vacuum polarization effect $\Delta_{\rm QED}$ can be neglected. Consequently, we have $\Delta_\parallel = \Delta_\perp = \Delta_{\rm pl}$ and the matrix element can be further simplied to $\Delta_{xx} = \Delta_{yy} =\Delta_{\rm pl}$, $\Delta_{xy}=\Delta_{yx}=0$. In our case, we consider ultralight ALPs with masses much smaller than eV. They satisfy the extremely relativistic approximation condition $E_a\gg m_a$. Therefore, the three wave numbers $\Delta_{a}, \Delta_{a\gamma}, \Delta_{{\rm pl}}$ can be defined as
\begin{eqnarray}
    \Delta_a \equiv -\frac{m_a^2}{2E_a},~\Delta_{a\gamma}\equiv \frac{g_{a\gamma\gamma}B}{2},~\Delta_{{\rm pl}}\equiv -\frac{\omega_{{\rm pl}}^2}{2E_a}\;,
\end{eqnarray}
where $\omega_{\rm pl}=\sqrt{4\pi\alpha n_e/m_e}\simeq 4\times 10^{-11}~\eV\sqrt{n_e/{\rm cm}^{-3}}$ is the effective plasma frequency of photons with $n_e$ being the free electrons density in medium.
Next, we use characteristic paramters to model the cluster environment. We assume that the cluster radius is $R=1~{\rm Mpc}$ and the coherence length of the turbulent magnetic field is $l_c=10~{\rm kpc}$~\cite{Govoni:2004as}. They lead to $k=R/l_c=100$ domains along the propagation path. Previous studies have shown that the magnetic field strength in galaxy clusters follows the radial distribution of the thermal electron distribution $n_e(r)$. They are given by~\cite{Meyer:2013pny}
\begin{eqnarray}
B_{\rm cluster}(r)=B^0_{\rm cluster}(n_e(r)/n_e^0)^\eta\;,~~n_e(r)=n_e^0(1+r/r_{\rm core})^{-3\beta/2}\;,
\end{eqnarray}
where the typical value of $\eta$ is in the range of $0.5\sim1$. We adopt $\eta=0.5$ and $\beta=2/3$ in our analysis. We choose the central magnetic field $B^0_{\rm cluster}=10~\mu{\rm G}$ and the electron number density $n_e^0=10^{-2}~{\rm cm}^{-3}$ for the most massive cluster with the typical core radius $r_\text{core}=100~\text{kpc}$~\cite{Vikhlinin:2005mp}. Since the magnetic $B(r)$ varies with radius, we determine the magnetic field value for each region by averaging over the radial integration.

Since the ALPs originate from PBH Hawking radiation, their momenta are assumed to have isotropical distribution and their spin is zero. Therefore, their initial density matrix is taken to be $\rho_0={\rm diag}(0,0,1)$. The final photon state consists of two linearly polarized pure states, resulting to $\rho_{\rm final}=\rho_{11}+\rho_{22}={\rm diag}(1,0,0)+{\rm diag}(0,1,0)$. Then, the conversion probability is given by the following result
\begin{eqnarray}
P_{a\to\gamma}^{\rm exgal}=\Tr[\rho_{\rm final}\mathcal{T}(\psi_k,\cdots,\psi_1)\rho_0\mathcal{T}^\dagger(\psi_k,\cdots,\psi_1)]\;,~\mathcal{T}(\psi_k,\cdots,\psi_1)=\prod_{k=1}\mathcal{T}_k=\prod_{k=1}e^{i\mathcal{M}_kl_c}\;.
\end{eqnarray}
To fully account for the effects of the turbulent magnetic field, we model the angle $\psi$ as a random variable uniformly distributed in the interval $[0,2\pi)$ and sample it independently in each domain. Assuming that the PBH is located at the center of the galaxy cluster, we perform a total of 1,000 Monte Carlo simulations. Each simulation involves global propagation using a set of randomly generated angles $\{\psi_k\}_{1\leq k\leq100}$. Subsequently, the median of the resulting conversion probabilities was taken as the final statistical result.

For calculating the probability of ALP-photon conversion in the Milky Way, we adopt the ``JF12'' model~\cite{Jansson:2012pc} to treat the galactic magnetic field structure. In JF12 model, the regular field is composed of disk, halo, and X-field terms. We use two simplification approaches in our analysis. The first simplification is to regard the Milky Way as a two-dimensional disk with $z=0$. This approach automatically eliminates the halo and X-field terms and leaves a purely planar field whose components lie along $\hat{\phi}$ and $\hat{r}$, corresponding to the molecular ring and spiral arm components of the disk field, respectively. The second simplification is motivated by the fact that galactic PBH distributions inherit DM profiles predicting a strong central concentration. We therefore consider only a single propagation path, extending from the Galactic center to the solar system and having a characteristic length of 8.2 kpc~\cite{Gravity:2019nxk}.
Along this path, the magnetic field strength decreases as $1/r$. We thus adopt a cell-like approach by splitting the path into eight segments as the number of spiral arms. In each segment, the field strength and orientation are treated as approximately uniform. This method permits the calculation to be simplified by rotating the angle $\psi$ to zero. As a result of this rotation, the ALPs  only couple to the photon polarization state along the $x_2$-direction. The $x_1$-polarized photon state remains completely decoupled. The propagation matrix $\mathcal{M}$ therefore simplifies to
\begin{eqnarray}
\mathcal{M}_0=\left(\begin{matrix}
    \Delta_\bot&0&0\\
        0&\Delta_\parallel&\Delta_{a\gamma}\\
        0&\Delta_{a\gamma}&\Delta_a
    \end{matrix}\right)\;.
\end{eqnarray}
Based on Ref.~\cite{Jansson:2012pc}, we calculate the averaged magnetic field in each domain by adopting a mean thermal electron density of $n_e=10^{-2}~{\rm cm}^{-3}$ in the Milky Way~\cite{Beck:2008ty,Cordes:2002wz}. We then follow the
method employed for galaxy clusters to calculate the conversion probability within the Milky Way $P_{a\to \gamma}^{\rm gal}$.

\begin{figure}
\centering
\includegraphics[width=0.49\linewidth]{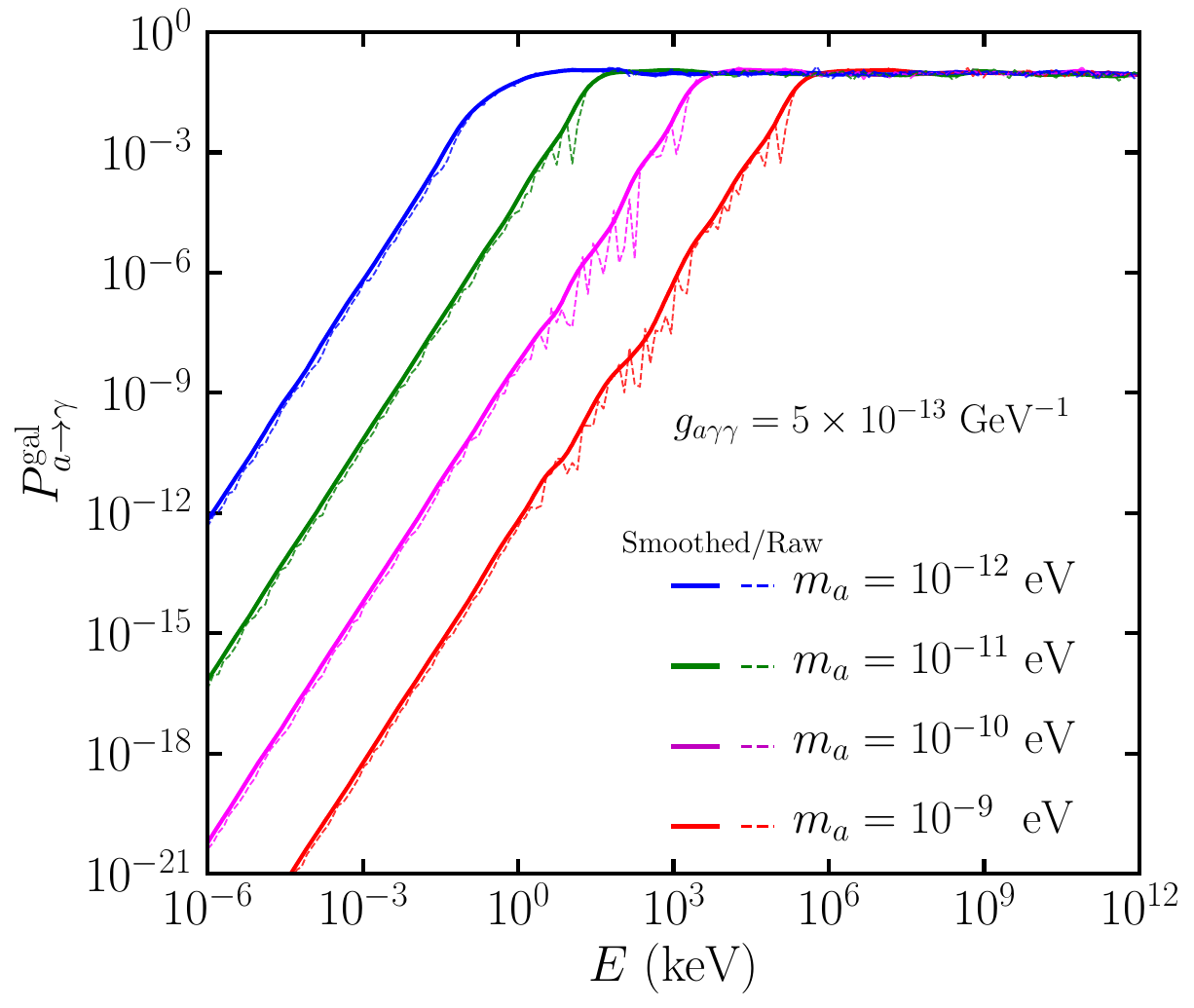}
\includegraphics[width=0.49\linewidth]{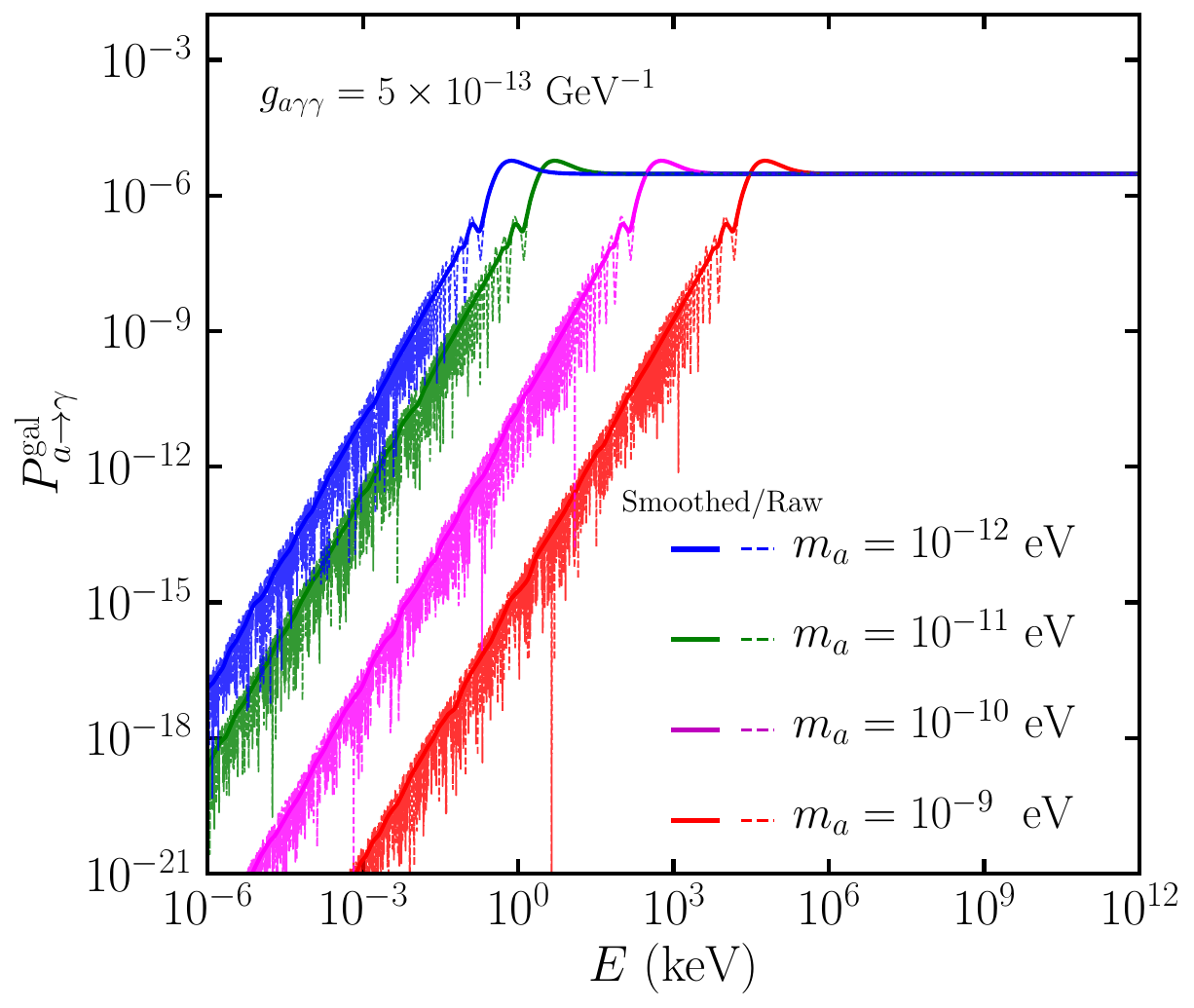}
\caption{Probability of ALP conversion into photon within the galaxy clusters $P_{a\to \gamma}^{\rm exgal}$ (left panel) and in the Milky Way $P_{a\to \gamma}^{\rm gal}$ (right panel). ALP mass is chosen as $m_a=10^{-12}$ eV (blue), $10^{-11}$ eV (green), $10^{-10}$ eV (purple) or $10^{-9}$ eV (red) and the ALP-photon coupling is fixed as $g_{a\gamma\gamma}=5\times 10^{-13}~\GeV^{-1}$.
We highlight the smoothed curves obtained through Gaussian filtering with solid lines.
}
\label{fig:conversion_probability}
\end{figure}

In Fig.~\ref{fig:conversion_probability}, we show the probability of ALP conversion into photon $P_{a\to \gamma}^{\rm exgal}$ within the galaxy clusters (left panel) and $P_{a\to \gamma}^{\rm gal}$ in the Milky Way (right panel), for four choices of ALP mass $m_a = 10^{-12}$ eV (blue line), $10^{-11}$ eV (green line), $10^{-10}$ eV (purple line), $10^{-9}$ eV (red line) and fixed coupling $g_{a\gamma\gamma} = 5\times 10^{-13}~{\rm GeV}^{-1}$. As the energy increases, one can see that the conversion probability increases and tends to be a constant. The stable probability in the galaxy clusters is four orders of magnitude larger than that in the Milky Way. We also smooth the results of conversion probability by Gaussian filtering to minimize the unstability caused by the oscillation in probability, as shown by the solid lines. The smoothed probability will be used in subsequent analyses.

The gamma-rays from ALP conversion then yield the following energy spectrum
\begin{equation}
\frac{d^2\Phi_{\rm sig}}{dEd\Omega}=\frac{d^2\Phi_{\rm gal}}{dEd\Omega}\times P_{a\to \gamma}^{\rm gal}(m_a,g_{a\gamma\gamma})+\frac{d^2\Phi_{\rm exgal}}{dEd\Omega}\times P_{a\to \gamma}^{\rm exgal}(m_a,g_{a\gamma\gamma})\;,
\end{equation}
where the superscripts ``gal'' and ``exgal'' of $P_{a\to \gamma}$ represent galactic region and extragalactic region, respectively.
The extragalactic regions can be categorized into the extragalactic background light (EBL)
and galaxy clusters. We consider both the conversion of PBH ALPs to photons within the Milky Way’s magnetic field and the conversion of extragalactic PBH ALPs in galaxy cluster magnetic fields. Due to the weak magnetic field strength of the EBL $B\sim {\rm nG}$~\cite{Planck:2015zrl,Pshirkov:2015tua}, we omit the conversion of PBH ALPs to the photon signal when passing through this region in subsequent calculations. In addition, PBHs can emit both ALPs and photons via Hawking evaporation. In the energy region that we consider, the PBH emitted photon flux can be significant and therefore plays as the major background in our case. We also use BlackHawk to generate the number density of photons from PBH evaporation and calculate the flux of PBH gamma-ray background.

Besides the gamma-rays from PBH evaporation, there exist other backgrounds in the universe. Current gamma-ray experiments such as Fermi-LAT~\cite{Fermi-LAT:2014ryh}, COMPTEL~\cite{Weidenspointner:2000aq}, and INTEGRAL~\cite{Tsygankov:2007jf} have provided a complete energy spectrum of the extragalactic photon background (EGB) in the range of ${\rm keV}-{\rm TeV}$. A fitted model is also provided
for the galactic gamma-ray background in the low-energy region~\cite{Bartels:2017dpb}. In the low-energy region of AMEGO, the gamma-ray background fluxes are parameterized as
\begin{align}
\left.\frac{d^2\Phi_{\rm exgal}^{\rm bkg}}{dE\,d\Omega}\right|_{150~\keV\le E\le 5~\MeV}
&= A_{\rm bkg}^{\rm exgal}\left(\frac{E}{1~\MeV}\right)^{-\alpha^{\rm exgal}}\;,\\
\left.\frac{d^2\Phi_{\rm gal}^{\rm bkg}}{dE\,d\Omega}\right|_{150~\keV\le E\le 5~\MeV}
&= A_{\rm bkg}^{\rm gal}\left(\frac{E}{1~\MeV}\right)^{-\alpha^{\rm gal}}
\exp\left[-\left(\frac{E}{E_c}\right)^{\gamma^{\rm gal}}\right]\;,
\end{align}
where the extragalactic component is modeled as a power law with
$A_{\rm bkg}^{\rm exgal}=0.004135~\MeV^{-1}\,\mathrm{cm}^{-2}\,\mathrm{s}^{-1}\,\mathrm{sr}^{-1}$
and $\alpha^{\rm exgal}=2.8956$~\cite{Ballesteros:2019exr}. For the galactic gamma-ray background, we adopt the model of Ref.~\cite{Bartels:2017dpb} with best-fit parameters
$A_{\rm bkg}^{\rm gal}=0.013~\MeV^{-1}\,\mathrm{cm}^{-2}\,\mathrm{s}^{-1}\,\mathrm{sr}^{-1}$,
$\alpha^{\rm gal}=1.8$, $E_c=20~\MeV$, and $\gamma^{\rm gal}=2$.
In the high-energy region of e-ASTROGAM and MAST, we adopt the following background model
\begin{align}
\left.\frac{d^2\Phi_{\rm exgal}^{\rm bkg}}{dE\,d\Omega}\right|_{100~\MeV\le E\le 3~\GeV}
&= A_{\rm bkg}^{\rm exgal}\left(\frac{E}{100~\MeV}\right)^{-\alpha^{\rm exgal}}
\exp\left[-\left(\frac{E}{E_c}\right)\right]\;,\\
\left.\frac{d^2\Phi_{\rm gal}^{\rm bkg}}{dE\,d\Omega}\right|_{100~\MeV\le E\le 3~\GeV}
&= A_{\rm bkg}^{\rm gal}\left(\frac{E}{100~\MeV}\right)^{-\alpha^{\rm gal}}
\exp\left[-\left(\frac{E}{E_c}\right)^{\gamma^{\rm gal}}\right]\;,
\end{align}
where the extragalactic model, a power law with an exponential cutoff, is taken from the Fermi-LAT Collaboration~\cite{Fermi-LAT:2014ryh} with
$A_{\rm bkg}^{\rm exgal}=1.48\times 10^{-7}~\MeV^{-1}\,\mathrm{cm}^{-2}\,\mathrm{s}^{-1}\,\mathrm{sr}^{-1}$,
$\alpha^{\rm exgal}=2.31$, and $E_c=362~\GeV$.
For the galactic component which dominates in this energy range, we use a fit to the pion-decay data from Ref.~\cite{Bartels:2017dpb}, with
$A_{\rm bkg}^{\rm gal}=0.00538~\MeV^{-1}\,\mathrm{cm}^{-2}\,\mathrm{s}^{-1}\,\mathrm{sr}^{-1}$,
$\alpha^{\rm gal}=3.32$, $E_c=45708~\MeV$, and $\gamma^{\rm gal}=-0.343$.

\begin{figure}[ht]
\begin{center}
\includegraphics[scale=1,width=0.9\linewidth]{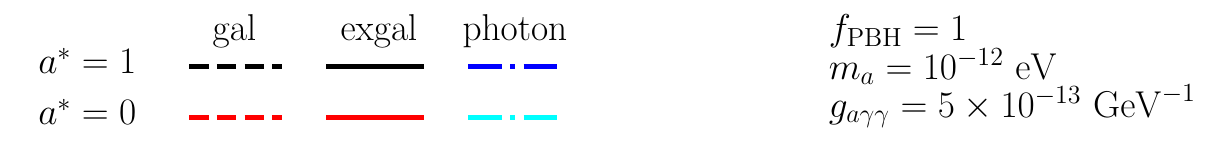}
\includegraphics[scale=1,width=0.49\linewidth]{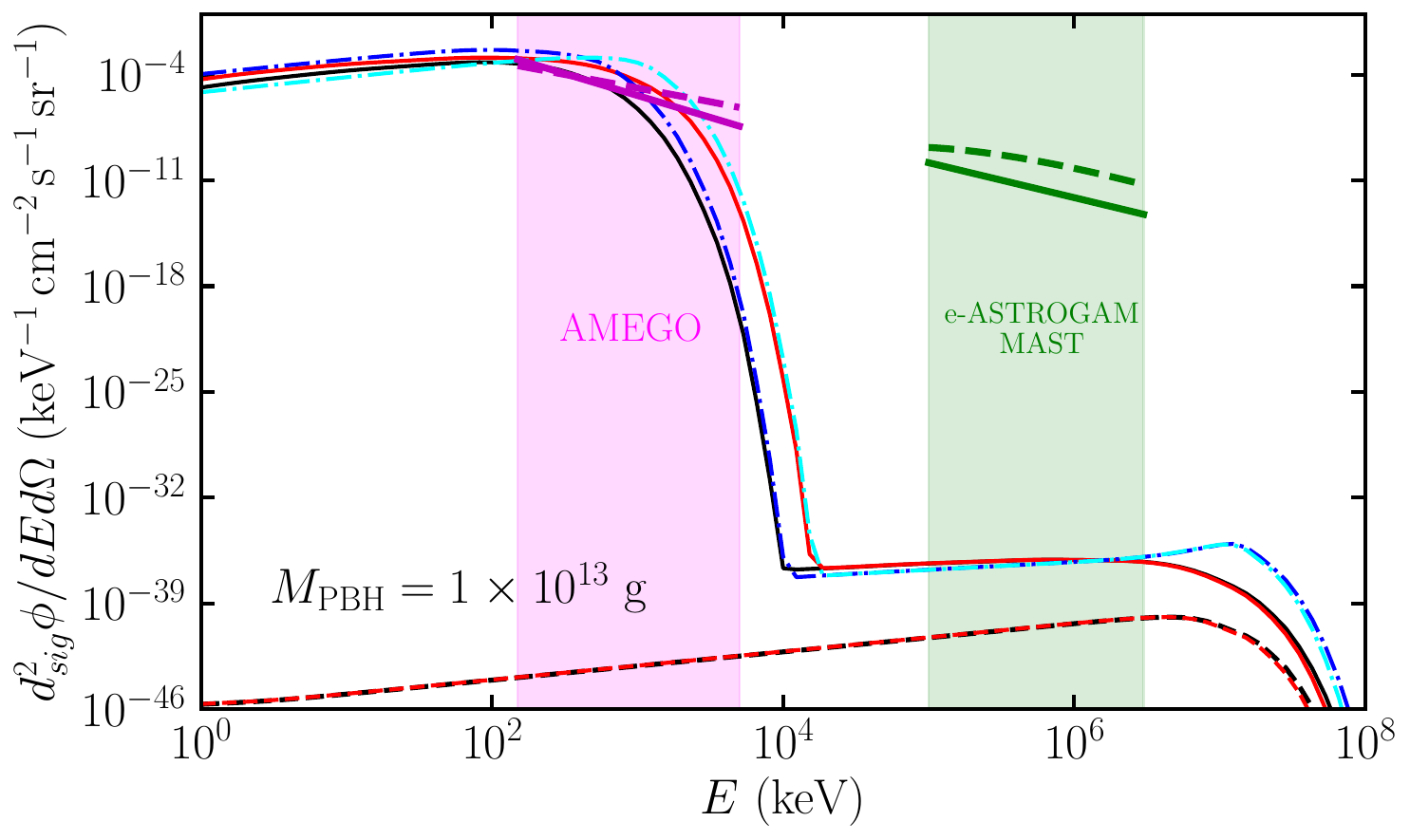}
\includegraphics[scale=1,width=0.49\linewidth]{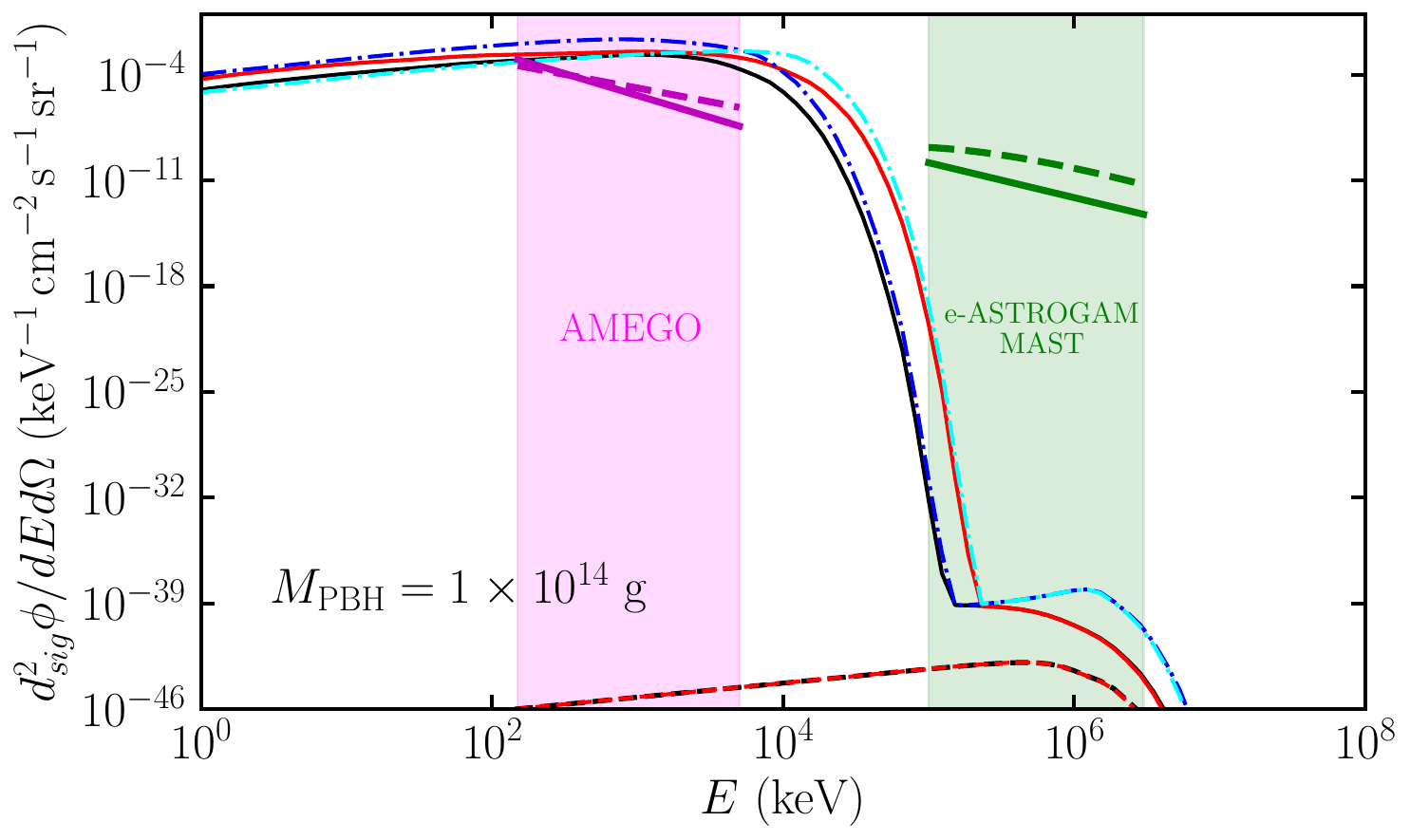}
\includegraphics[scale=1,width=0.49\linewidth]{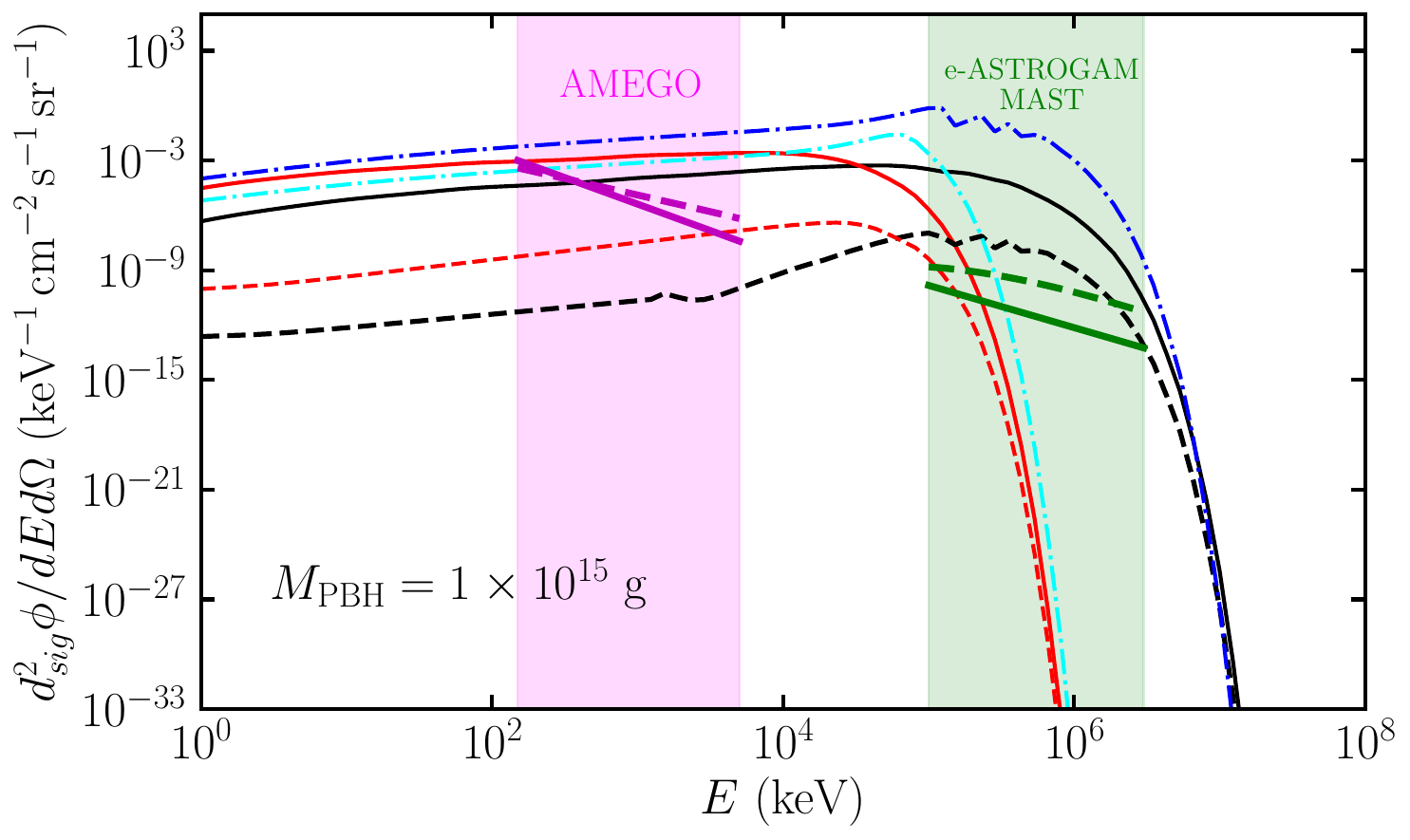}
\includegraphics[scale=1,width=0.49\linewidth]{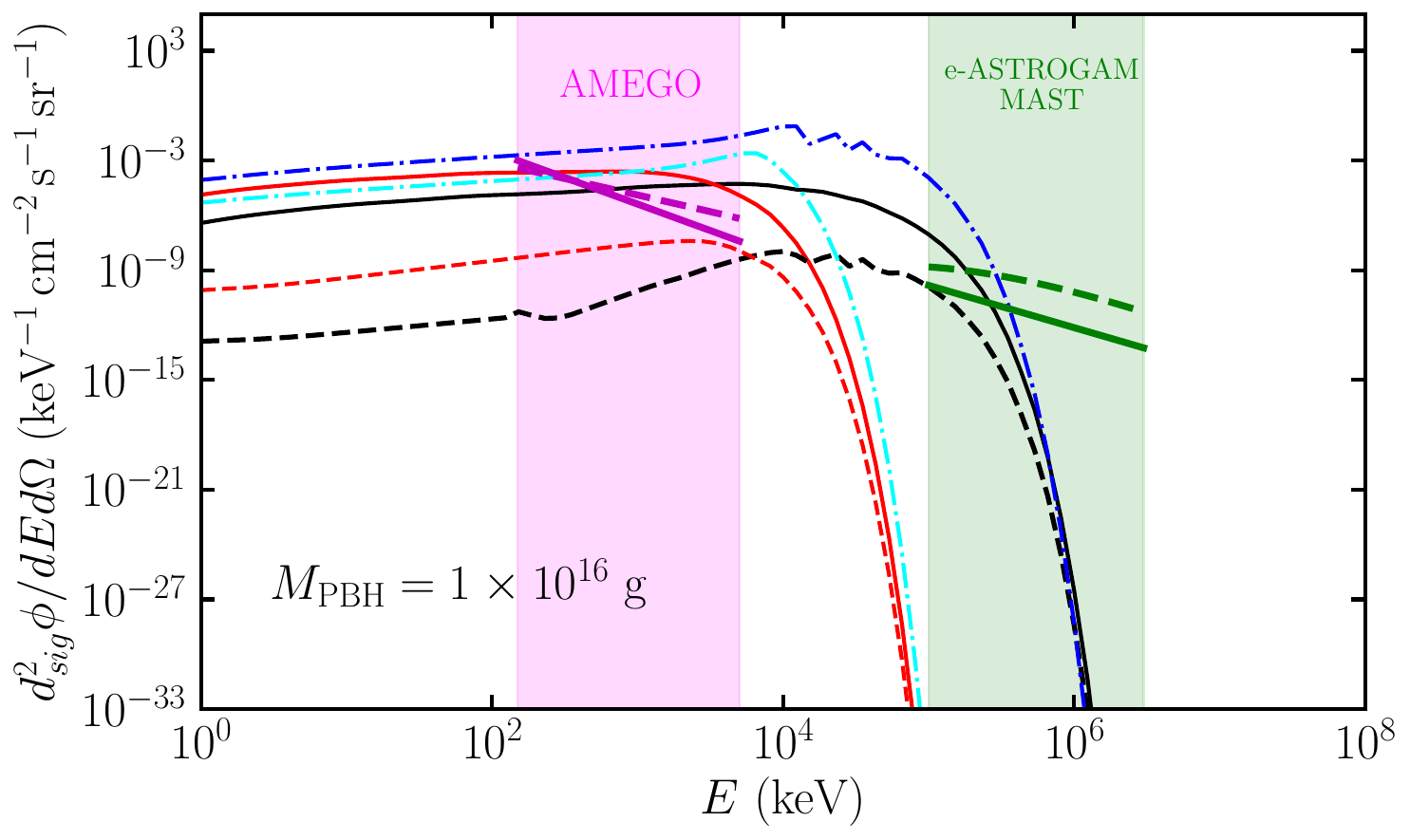}
\end{center}
\caption{
Differential gamma-ray flux from PBH emitted ALPs after propagating through the extragalactic (solid lines) or galactic (dashed lines) background magnetic fields for $M_{\rm PBH}=1\times10^{13}~\rm g$ (top left panel), $M_{{\rm PBH}}=1\times10^{14}~\rm g$ (top right panel), $M_{\rm PBH}=1\times10^{15}~\rm g$ (bottom left panel), $M_{{\rm PBH}}=1\times10^{16}~\rm g$ (bottom right panel), with $f_{\rm PBH}=1$, $m_a=10^{-12}~{\rm eV}$ and $g_{a\gamma\gamma}=5\times 10^{-13}~{\rm GeV}^{-1}$. The two cases of PBH spin are considered as $a^*=1$ (black lines) and $a^*=0$ (red lines). The total PBH gamma-ray backgrounds for the two cases are also shown by dash-dotted curves (blue for $a^*=1$ and cyan for $a^*=0$). The energy windows of AMEGO ($150~{\rm keV}-5~{\rm MeV}$, purple) and e-ASTROGAM/MAST ($100~{\rm MeV}-3~{\rm GeV}$, green) are also shown for reference. We also show the EGB using solid line and the galactic gamma-ray background using dashed line in individual energy window.
}
\label{fig:spectrum}
\end{figure}

In Fig.~\ref{fig:spectrum}, we show the differential gamma-ray flux from PBH-emitted ALPs after propagating through the extragalactic (solid lines) or galactic (dashed lines) background magnetic fields for $M_{\rm PBH}=1\times10^{13}~\rm g$ (top left panel), $M_{{\rm PBH}}=1\times10^{14}~\rm g$ (top right panel), $M_{\rm PBH}=1\times10^{15}~\rm g$ (bottom left panel), $M_{{\rm PBH}}=1\times10^{16}~\rm g$ (bottom right panel), with $f_{\rm PBH}=1$, $m_a=10^{-12}~{\rm eV}$ and $g_{a\gamma\gamma}=5\times 10^{-13}~{\rm GeV}^{-1}$.
The two cases of PBH spin are considered as $a^*=1$ (black lines) and $a^*=0$ (red lines). The total PBH gamma-ray backgrounds for the two cases are also shown by dash-dotted curves (blue for $a^*=1$ and cyan for $a^*=0$). They are comparable to the gamma-ray signal converted by PBH ALPs.
The energy windows for AMEGO ($150~{\rm keV}-5~{\rm MeV}$, light purple) and e-ASTROGAM/MAST ($100~{\rm MeV}-3~{\rm GeV}$, light green) are also shown for reference.
In each energy window, we display the EGB using solid line and the galactic gamma-ray background using dashed line. In the plots on the top, one can see that the extragalactic PBH contributions before the memory-burden phase are dominant in the AMEGO energy window. The extragalactic PBH contributions are typically suppressed in the memory-burden phase, arising in the e-ASTROGAM/MAST window. The galactic PBH components are further suppressed by several orders of magnitude due to the conversion probability. As a result,
the AMEGO experiment with lower energy window is suitable to
probe the evaporation-conversion signals from both high-mass and low-mass PBHs. On the other hand, higher-mass PBHs shift the dominant extragalactic contribution to e-ASTROGAM/MAST energy region. These experiments are thus more sensitive to the signal from higher-mass PBHs. However, as seen from the bottom panels, once the PBH mass exceeds $\sim 10^{15}$ g, e-ASTROGAM and MAST soon become insensitive for $a^\ast=0$ PBHs. This is because the emission energy decreases and no longer falls within their detectable window.
The nonzero spin of PBHs can defer this insensitivity.

%%%%%%%%%%%%%%%%%%%%%%%%%%%%%%%%%%%%
\section{Prospective detection of memory-burdened PBH and ALP-photon coupling}
\label{sec:Sensitivity}
%%%%%%%%%%%%%%%%%%%%%%%%%%%%%%%%%%%%

Based on the energy spectra of the gamma-ray signal from PBH ALP and backgrounds, we adopt the Fisher forecasting approach~\cite{Edwards:2017mnf} to numerically assess the detection prospects for memory-burdened PBHs and light ALPs. Compared with the Maximum Likelihood Ratio (MLR) method, Fisher forecasting offers the advantages of requiring fewer parameters and allowing a simpler model. In order to quantify how sensitive the observed data are to the chosen parameters, the main idea of this approach is to employ the Fisher information matrix calculated from the first and second derivatives of the signal spectrum function. This method also has excellent local approximation properties. Mathematically, the Fisher information matrix is an \(n\)-dimensional symmetric positive-definite matrix, defined as the second derivative of the log-likelihood function \(\ln{\mathcal{L}(\vec\theta\vert\mathcal{D})}\) with respect to the \(n\)-dimensional parameter vector \(\vec{\theta}\) for a given data set \(\mathcal{D}\). The eigenvalues of the Fisher information matrix quantify the sensitivity of parameter estimation along orthogonal directions in parameter space. Ref.~\cite{Bartels:2017dpb} presents a generalized signal-to-noise ratio (SNR) formulation of the Fisher matrix
\begin{eqnarray}
\mathcal{I}_{ij}=\int_{E_{{\rm min}}}^{E_{\rm max}}dE\int_{{\rm ROI}}d\Omega\mathcal{E}(E)\frac{\partial_i\phi(E,\Omega)\partial_j\phi(E,\Omega)}{\phi(E,\Omega\vert\vec\theta=1)}\;,
\end{eqnarray}
where $\phi(E,\Omega)=d^2\Phi/dEd\Omega$ is the combined contribution of the signal and background fluxes with $\phi=\phi_{{\rm sig}}+\phi_{{\rm bkg}}$, and $\vec{\theta}=1$ means that all parameters of the function $\phi(E,\Omega)$ are set to 1. $\mathcal{E}(E)=T_{{\rm obs}}A_{{\rm eff}}(E)$ represents the instrumental exposure for gamma-ray detection, where $T_{{\rm obs}}$ is the observation time and $A_{{\rm eff}}(E)$ is the energy-dependent effective area. In this work, $T_{{\rm obs}}$ is set to 1 year, and $A_{{\rm eff}}(E)$ is taken from the three detectors AMEGO~\cite{Kierans:2020otl}, e-ASTROGAM~\cite{e-ASTROGAM:2016bph}, and MAST~\cite{Dzhatdoev:2019kay}.

The above Fisher information matrix is used to calculate the exclusion limits on a specific set of parameters $\vec\theta$.
In practice, the nuisance parameters such as background parameters are typically profiled out to derive the ``profiled Fisher information matrix''. For instance, to project the limits in the $f_{\rm{PBH}}$ versus $M_{\rm PBH}$ parameter space, our case incorporates two types of parameters for a fixed $M_{\rm PBH}$: $f_{\rm PBH}$ as the only parameter of interest (PoI), and the nuisance parameters $A_{\rm{bkg}}^{\rm{exgal}}$, $\alpha^{\rm exgal}$, $A_{\rm{bkg}}^{\rm gal}$, $\alpha^{\rm gal}$, $E_c$ and $\gamma^{\rm gal}$ for AMEGO, or $A_{\rm{bkg}}^{\rm gal}$, $\alpha^{\rm gal}$, $E_c$ and $\gamma^{\rm gal}$ for e-ASTROGAM and MAST. Note that, since the extragalactic background is at least one order of magnitude smaller than the galactic background in the high-energy region, we neglect extragalactic components in the Fisher forecasting for e-ASTROGAM and MAST. The Fisher matrix is then partitioned into a block form as follows
\begin{eqnarray}
    \mathcal{I}_{ij}=
    \begin{pmatrix}
        \mathcal{I}_A & \mathcal{I}_C^T \\
        \mathcal{I}_C & \mathcal{I}_B
    \end{pmatrix}\;,
\end{eqnarray}
where the Fisher matrix element $\mathcal{I}_A=\mathcal{I}_{00}$ quantifies the sensitivity of the signal to variations in the PBH fraction $f_{{\rm PBH}}$, and the remaining components $\mathcal{I}_C$, $\mathcal{I}_C^T$, and $\mathcal{I}_B$, are relevant for background parameters. Using the Schur complement lemma, one can define the profiled Fisher information matrix as $\widetilde{\mathcal{I}}_A=\mathcal{I}_A-\mathcal{I}_C^T~\mathcal{I}_B^{-1}\mathcal{I}_C$. Thus, $\widetilde{\mathcal{I}}_A$ preserves the statistical information of the PoI while eliminating covariance interference introduced by nuisance parameters. The inverse of each diagonal element of ${\widetilde{\mathcal{I}}_A}$ results in the upper limit for the corresponding parameter~\cite{Bartels:2017dpb}
\begin{eqnarray}
\theta_1^{\mathrm{UL}} = f_{\mathrm{PBH}}^{\mathrm{UL}} = Z(p) \sqrt{\widetilde{\mathcal{I}}^{-1}_{A,11}}\;,
\end{eqnarray}
where $Z(p)$ is the cumulative distribution function. We consider the upper limit of $f_{\rm PBH}$ at the 95\% confidence level (C.L.) with $Z(0.05)=1.645$.

\begin{figure}[htbp]
\begin{center}
\includegraphics[scale=1,width=0.49\linewidth]{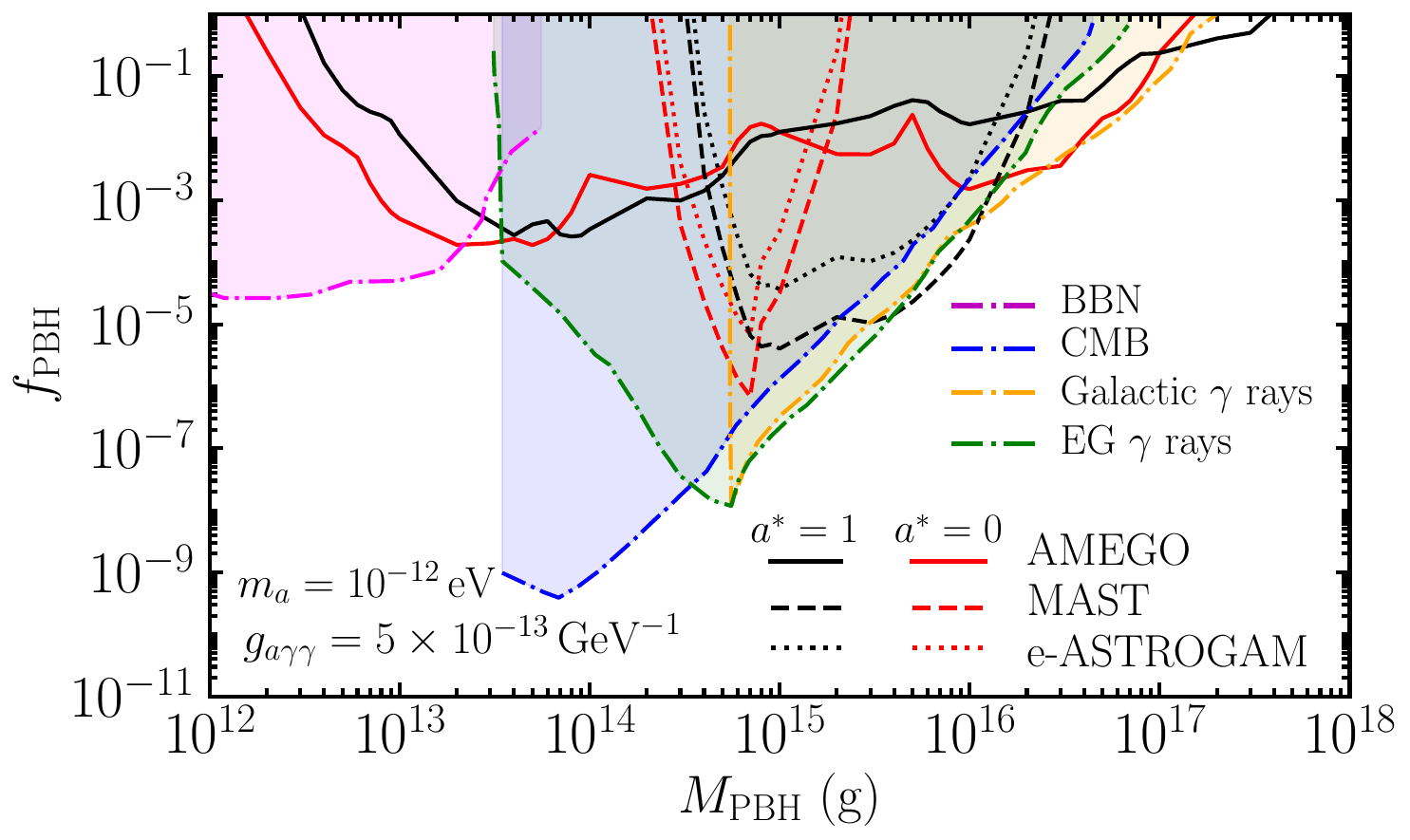}
\includegraphics[scale=1,width=0.49\linewidth]{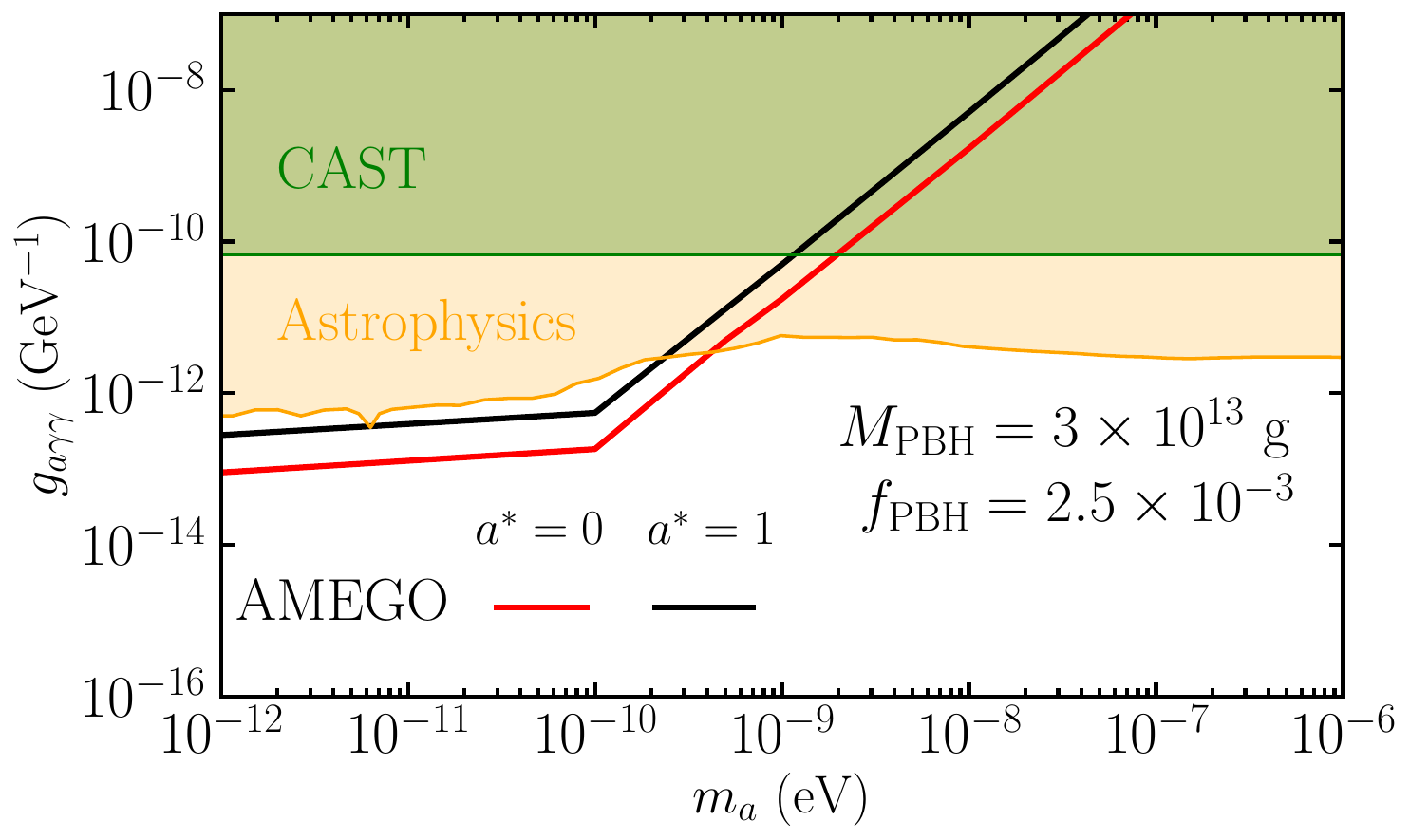}
\end{center}
\caption{Left: Projected 95\% C.L. upper limits on the PBH DM fraction $f_{\rm PBH}$ from the evaporation-conversion scenario for AMEGO (solid lines), e-ASTROGAM (dotted lines) and MAST (dashed lines), with fixed ALP-photon coupling $g_{a\gamma\gamma} = 5\times10^{-13}~\GeV^{-1}$ and ALP mass $m_a = 10^{-12}~\eV$.
The two cases of PBH spin are considered as $a^\ast = 1$ (black lines) and $a^\ast = 0$ (red lines). Other constraints for $k=2$~\cite{Thoss:2024hsr} are also shown by dash-dotted lines and shaded regions.
Right: Projected 95\% C.L. upper limits on ALP-photon coupling $g_{a\gamma\gamma}$ from AMEGO experiment with $M_{\rm PBH}=3\times 10^{13}$ g and $f_{\rm PBH}=2.5\times 10^{-3}$. The shaded regions show parameter space excluded by CAST~\cite{CAST:2017uph} and astrophysical observations~\cite{AxionLimits}.
}
\label{fig:upper_limit}
\end{figure}

In the left panel of Fig.~\ref{fig:upper_limit}, we display the projected 95\% C.L. upper limits on the PBH DM fraction $f_{\rm PBH}$ from the evaporation-conversion scenario for AMEGO (solid lines), e-ASTROGAM (dotted lines) and MAST (dashed lines), with fixed ALP-photon coupling $g_{a\gamma\gamma} = 5\times10^{-13}~\GeV^{-1}$ and ALP mass $m_a = 10^{-12}~\eV$.
The PBH spins are taken as $a^\ast = 1$ (black lines) and $a^\ast = 0$ (red lines). The power of PBH entropy in memory-burden phase is also fixed as $k=1$. It is worth emphasizing that our results remain essentially unchanged whether we choose $k=1$ or $k=2$. For comparison, we also present other existing constraints obtained with $k=2$~\cite{Thoss:2024hsr} in dash-dotted lines and shaded regions.
AMEGO with lower detectable energy window is sensitive to very broad range of PBH mass from $10^{12}$ g to $10^{17}$ g. The e-ASTROGAM and MAST experiments with their high-energy coverage can probe a narrower PBH mass range from $2\times 10^{14}$ g to $2\times 10^{15}$ g for $a^\ast=0$. For PBH spin $a^\ast=1$, the detectable PBH mass can be extended to $3\times 10^{16}$ g. Due to the smaller gamma-ray background in e-ASTROGAM/MAST energy window and their larger effective
area, e-ASTROGAM and MAST can reach $f_{\rm PBH}$ values two orders of magnitude lower than AMEGO.

Next, we reverse the Fisher forecasting procedure for $f_{\rm PBH}$-$M_{\rm PBH}$ to obtain the reachable limits on ALP-photon coupling $g_{a\gamma\gamma}$ for specific PBH mass. For illustration, we take $M_{\rm PBH}=3\times 10^{13}$ g and $f_{\rm PBH}=2.5\times 10^{-3}$ which are allowed by cosmological constraints~\cite{Thoss:2024hsr}. The right panel of Fig.~\ref{fig:upper_limit} shows the 95\% C.L. upper limits on $g_{a\gamma\gamma}$ versus $m_a$ from AMEGO. It turns out that AMEGO can probe ALP-photon coupling down to $g_{a\gamma\gamma}\sim 10^{-13}~{\rm GeV}^{-1}$ for $m_a<10^{-10}~{\rm eV}$. This improves current astrophysical constraints by one order of magnitude. The limit becomes significantly weaker for $m_a>10^{-10}~{\rm eV}$, due to rapid decline of conversion probability within the integral region.

%%%%%%%%%%%%%%%%%%%%%%%%
\section{Conclusion}
\label{sec:Con}
%%%%%%%%%%%%%%%%%%%%%%%%

The ALP as light bosonic particle and PBH as macroscopic object are two representative DM candidates. In this work, based on the recently proposed memory-burden effect, we systematically explore the gamma-ray production mechanisms via memory-burdened PBH evaporation and ALP-photon coupling $g_{a\gamma\gamma}$. Furthermore, we evaluate the detection prospects of this evaporation-conversion signal in next-generation satellite telescopes, including AMEGO, e-ASTROGAM and MAST.

We consider the evaporation-conversion scenario in which light ALPs are emitted by memory-burdened PBHs and are converted into photons in the presence of magnetic field in the universe. The PBH
evaporation progress is divided into the semiclassical phase of Hawking radiation and the memory-burden phase of quantum effect. The time evolution of PBH mass and the gamma-ray flux after the conversion of PBH-emitted ALP are then evaluated by taking into account the memory-burden induced PBH entropy suppression.
Applying the Fisher forecasting method, we calculate the gamma-ray energy spectra for this scenario and derive the projected sensitivity for PBH fraction $f_{\rm PBH}$ and ALP-photon coupling $g_{a\gamma\gamma}$. Our main conclusions are summarized as follows.
\begin{itemize}
\item Before the memory-burden phase, the gamma-ray flux from the light extragalactic PBHs are dominant in the AMEGO energy window. The memory-burden effect suppresses the gamma-ray flux in e-ASTROGAM/MAST energy window unless for higher-mass PBHs. For the PBHs with mass larger than $\sim 10^{15}$ g, e-ASTROGAM and MAST become insensitive for $a^\ast=0$ PBHs and the nonzero spin of PBHs can defer this insensitivity.
\item AMEGO with lower detectable energy window is sensitive to very broad range of PBH mass from $10^{12}$ g to $10^{17}$ g. The e-ASTROGAM and MAST experiments with their high-energy coverage can probe a narrower PBH mass range from $2\times 10^{14}$ g to $2\times 10^{15}$ g for $a^\ast=0$. For PBH spin $a^\ast=1$, the detectable PBH mass can be extended to $3\times 10^{16}$ g. e-ASTROGAM and MAST can reach $f_{\rm PBH}$ values two orders of magnitude lower than AMEGO.
\item AMEGO can probe ALP-photon coupling down to $g_{a\gamma\gamma}\sim 10^{-13}~{\rm GeV}^{-1}$ for $m_a<10^{-10}~{\rm eV}$. This improves current astrophysical constraints by one order of magnitude.
\end{itemize}

%###################################################################
%%%%%%%%%%%%%%%%%%%%%%%%
\acknowledgments
%%%%%%%%%%%%%%%%%%%%%%%%
We would like to thank Jiajun Liao, Rui-Jia Zhang and Tong Zhu for useful discussions.
T.~L. is supported by the National Natural Science Foundation of China (Grant No.~12375096).

%%%%%%%%%%%%%%%%%%%%%%%%%%%%%%%%%%%%%%%%%%%%%%
%%%%%%%%%%%%%%%%%%%%%%%%%%%%%%%%%%%%%%%%%%%%%%
%%%%%%%%%%%%%%%%%%%%%%%%%%%%%%%%%%%%%
\bibliography{refs}

\begin{thebibliography}{99}%
\makeatletter
\providecommand \@ifxundefined [1]{%
 \@ifx{#1\undefined}
}%
\providecommand \@ifnum [1]{%
 \ifnum #1\expandafter \@firstoftwo
 \else \expandafter \@secondoftwo
 \fi
}%
\providecommand \@ifx [1]{%
 \ifx #1\expandafter \@firstoftwo
 \else \expandafter \@secondoftwo
 \fi
}%
\providecommand \natexlab [1]{#1}%
\providecommand \enquote  [1]{``#1''}%
\providecommand \bibnamefont  [1]{#1}%
\providecommand \bibfnamefont [1]{#1}%
\providecommand \citenamefont [1]{#1}%
\providecommand \href@noop [0]{\@secondoftwo}%
\providecommand \href [0]{\begingroup \@sanitize@url \@href}%
\providecommand \@href[1]{\@@startlink{#1}\@@href}%
\providecommand \@@href[1]{\endgroup#1\@@endlink}%
\providecommand \@sanitize@url [0]{\catcode `\\12\catcode `\$12\catcode
  `\&12\catcode `\#12\catcode `\^12\catcode `\_12\catcode `\%12\relax}%
\providecommand \@@startlink[1]{}%
\providecommand \@@endlink[0]{}%
\providecommand \url  [0]{\begingroup\@sanitize@url \@url }%
\providecommand \@url [1]{\endgroup\@href {#1}{\urlprefix }}%
\providecommand \urlprefix  [0]{URL }%
\providecommand \Eprint [0]{\href }%
\providecommand \doibase [0]{http://dx.doi.org/}%
\providecommand \selectlanguage [0]{\@gobble}%
\providecommand \bibinfo  [0]{\@secondoftwo}%
\providecommand \bibfield  [0]{\@secondoftwo}%
\providecommand \translation [1]{[#1]}%
\providecommand \BibitemOpen [0]{}%
\providecommand \bibitemStop [0]{}%
\providecommand \bibitemNoStop [0]{.\EOS\space}%
\providecommand \EOS [0]{\spacefactor3000\relax}%
\providecommand \BibitemShut  [1]{\csname bibitem#1\endcsname}%
\let\auto@bib@innerbib\@empty
%</preamble>
\bibitem [{\citenamefont {Zel'dovich}\ and\ \citenamefont
  {Novikov}(1967)}]{Zeldovich:1967lct}%
  \BibitemOpen
  \bibfield  {author} {\bibinfo {author} {\bibfnamefont {Y.~B.}\ \bibnamefont
  {Zel'dovich}}\ and\ \bibinfo {author} {\bibfnamefont {I.~D.}\ \bibnamefont
  {Novikov}},\ }\href@noop {} {\bibfield  {journal} {\bibinfo  {journal} {Sov.
  Astron.}\ }\textbf {\bibinfo {volume} {10}},\ \bibinfo {pages} {602}
  (\bibinfo {year} {1967})}\BibitemShut {NoStop}%
\bibitem [{\citenamefont {Carr}\ and\ \citenamefont
  {Hawking}(1974)}]{Carr:1974nx}%
  \BibitemOpen
  \bibfield  {author} {\bibinfo {author} {\bibfnamefont {B.~J.}\ \bibnamefont
  {Carr}}\ and\ \bibinfo {author} {\bibfnamefont {S.~W.}\ \bibnamefont
  {Hawking}},\ }\href {\doibase 10.1093/mnras/168.2.399} {\bibfield  {journal}
  {\bibinfo  {journal} {Mon. Not. Roy. Astron. Soc.}\ }\textbf {\bibinfo
  {volume} {168}},\ \bibinfo {pages} {399} (\bibinfo {year}
  {1974})}\BibitemShut {NoStop}%
\bibitem [{\citenamefont {Carr}(1975)}]{Carr:1975qj}%
  \BibitemOpen
  \bibfield  {author} {\bibinfo {author} {\bibfnamefont {B.~J.}\ \bibnamefont
  {Carr}},\ }\href {\doibase 10.1086/153853} {\bibfield  {journal} {\bibinfo
  {journal} {Astrophys. J.}\ }\textbf {\bibinfo {volume} {201}},\ \bibinfo
  {pages} {1} (\bibinfo {year} {1975})}\BibitemShut {NoStop}%
\bibitem [{\citenamefont {Khlopov}(2010)}]{Khlopov:2008qy}%
  \BibitemOpen
  \bibfield  {author} {\bibinfo {author} {\bibfnamefont {M.~Y.}\ \bibnamefont
  {Khlopov}},\ }\href {\doibase 10.1088/1674-4527/10/6/001} {\bibfield
  {journal} {\bibinfo  {journal} {Res. Astron. Astrophys.}\ }\textbf {\bibinfo
  {volume} {10}},\ \bibinfo {pages} {495} (\bibinfo {year} {2010})},\ \Eprint
  {http://arxiv.org/abs/0801.0116} {arXiv:0801.0116 [astro-ph]} \BibitemShut
  {NoStop}%
\bibitem [{\citenamefont {Barrau}\ \emph {et~al.}(2004)\citenamefont {Barrau},
  \citenamefont {Blais}, \citenamefont {Boudoul},\ and\ \citenamefont
  {Polarski}}]{Barrau:2003xp}%
  \BibitemOpen
  \bibfield  {author} {\bibinfo {author} {\bibfnamefont {A.}~\bibnamefont
  {Barrau}}, \bibinfo {author} {\bibfnamefont {D.}~\bibnamefont {Blais}},
  \bibinfo {author} {\bibfnamefont {G.}~\bibnamefont {Boudoul}}, \ and\
  \bibinfo {author} {\bibfnamefont {D.}~\bibnamefont {Polarski}},\ }\href
  {\doibase 10.1002/andp.200310067} {\bibfield  {journal} {\bibinfo  {journal}
  {Annalen Phys.}\ }\textbf {\bibinfo {volume} {13}},\ \bibinfo {pages} {115}
  (\bibinfo {year} {2004})},\ \Eprint {http://arxiv.org/abs/astro-ph/0303330}
  {arXiv:astro-ph/0303330} \BibitemShut {NoStop}%
\bibitem [{\citenamefont {Carr}\ and\ \citenamefont
  {Kuhnel}(2022)}]{Carr:2021bzv}%
  \BibitemOpen
  \bibfield  {author} {\bibinfo {author} {\bibfnamefont {B.}~\bibnamefont
  {Carr}}\ and\ \bibinfo {author} {\bibfnamefont {F.}~\bibnamefont {Kuhnel}},\
  }\href {\doibase 10.21468/SciPostPhysLectNotes.48} {\bibfield  {journal}
  {\bibinfo  {journal} {SciPost Phys. Lect. Notes}\ }\textbf {\bibinfo {volume}
  {48}},\ \bibinfo {pages} {1} (\bibinfo {year} {2022})},\ \Eprint
  {http://arxiv.org/abs/2110.02821} {arXiv:2110.02821 [astro-ph.CO]}
  \BibitemShut {NoStop}%
\bibitem [{\citenamefont {Hawking}(1974)}]{Hawking:1974rv}%
  \BibitemOpen
  \bibfield  {author} {\bibinfo {author} {\bibfnamefont {S.~W.}\ \bibnamefont
  {Hawking}},\ }\href {\doibase 10.1038/248030a0} {\bibfield  {journal}
  {\bibinfo  {journal} {Nature}\ }\textbf {\bibinfo {volume} {248}},\ \bibinfo
  {pages} {30} (\bibinfo {year} {1974})}\BibitemShut {NoStop}%
\bibitem [{\citenamefont {Hawking}(1975)}]{Hawking:1975vcx}%
  \BibitemOpen
  \bibfield  {author} {\bibinfo {author} {\bibfnamefont {S.~W.}\ \bibnamefont
  {Hawking}},\ }\href {\doibase 10.1007/BF02345020} {\bibfield  {journal}
  {\bibinfo  {journal} {Commun. Math. Phys.}\ }\textbf {\bibinfo {volume}
  {43}},\ \bibinfo {pages} {199} (\bibinfo {year} {1975})},\ \bibinfo {note}
  {[Erratum: Commun.Math.Phys. 46, 206 (1976)]}\BibitemShut {NoStop}%
\bibitem [{\citenamefont {Carr}\ \emph {et~al.}(2021)\citenamefont {Carr},
  \citenamefont {Kohri}, \citenamefont {Sendouda},\ and\ \citenamefont
  {Yokoyama}}]{Carr:2020gox}%
  \BibitemOpen
  \bibfield  {author} {\bibinfo {author} {\bibfnamefont {B.}~\bibnamefont
  {Carr}}, \bibinfo {author} {\bibfnamefont {K.}~\bibnamefont {Kohri}},
  \bibinfo {author} {\bibfnamefont {Y.}~\bibnamefont {Sendouda}}, \ and\
  \bibinfo {author} {\bibfnamefont {J.}~\bibnamefont {Yokoyama}},\ }\href
  {\doibase 10.1088/1361-6633/ac1e31} {\bibfield  {journal} {\bibinfo
  {journal} {Rept. Prog. Phys.}\ }\textbf {\bibinfo {volume} {84}},\ \bibinfo
  {pages} {116902} (\bibinfo {year} {2021})},\ \Eprint
  {http://arxiv.org/abs/2002.12778} {arXiv:2002.12778 [astro-ph.CO]}
  \BibitemShut {NoStop}%
\bibitem [{\citenamefont {Dvali}(2018)}]{Dvali:2018xpy}%
  \BibitemOpen
  \bibfield  {author} {\bibinfo {author} {\bibfnamefont {G.}~\bibnamefont
  {Dvali}},\ }\href@noop {} {\  (\bibinfo {year} {2018})},\ \Eprint
  {http://arxiv.org/abs/1810.02336} {arXiv:1810.02336 [hep-th]} \BibitemShut
  {NoStop}%
\bibitem [{\citenamefont {Dvali}\ \emph {et~al.}(2020)\citenamefont {Dvali},
  \citenamefont {Eisemann}, \citenamefont {Michel},\ and\ \citenamefont
  {Zell}}]{Dvali:2020wft}%
  \BibitemOpen
  \bibfield  {author} {\bibinfo {author} {\bibfnamefont {G.}~\bibnamefont
  {Dvali}}, \bibinfo {author} {\bibfnamefont {L.}~\bibnamefont {Eisemann}},
  \bibinfo {author} {\bibfnamefont {M.}~\bibnamefont {Michel}}, \ and\ \bibinfo
  {author} {\bibfnamefont {S.}~\bibnamefont {Zell}},\ }\href {\doibase
  10.1103/PhysRevD.102.103523} {\bibfield  {journal} {\bibinfo  {journal}
  {Phys. Rev. D}\ }\textbf {\bibinfo {volume} {102}},\ \bibinfo {pages}
  {103523} (\bibinfo {year} {2020})},\ \Eprint
  {http://arxiv.org/abs/2006.00011} {arXiv:2006.00011 [hep-th]} \BibitemShut
  {NoStop}%
\bibitem [{\citenamefont {Dvali}\ \emph {et~al.}(2024)\citenamefont {Dvali},
  \citenamefont {Valbuena-Berm{\'u}dez},\ and\ \citenamefont
  {Zantedeschi}}]{Dvali:2024hsb}%
  \BibitemOpen
  \bibfield  {author} {\bibinfo {author} {\bibfnamefont {G.}~\bibnamefont
  {Dvali}}, \bibinfo {author} {\bibfnamefont {J.~S.}\ \bibnamefont
  {Valbuena-Berm{\'u}dez}}, \ and\ \bibinfo {author} {\bibfnamefont
  {M.}~\bibnamefont {Zantedeschi}},\ }\href {\doibase
  10.1103/PhysRevD.110.056029} {\bibfield  {journal} {\bibinfo  {journal}
  {Phys. Rev. D}\ }\textbf {\bibinfo {volume} {110}},\ \bibinfo {pages}
  {056029} (\bibinfo {year} {2024})},\ \Eprint
  {http://arxiv.org/abs/2405.13117} {arXiv:2405.13117 [hep-th]} \BibitemShut
  {NoStop}%
\bibitem [{\citenamefont {Thoss}\ \emph {et~al.}(2024)\citenamefont {Thoss},
  \citenamefont {Burkert},\ and\ \citenamefont {Kohri}}]{Thoss:2024hsr}%
  \BibitemOpen
  \bibfield  {author} {\bibinfo {author} {\bibfnamefont {V.}~\bibnamefont
  {Thoss}}, \bibinfo {author} {\bibfnamefont {A.}~\bibnamefont {Burkert}}, \
  and\ \bibinfo {author} {\bibfnamefont {K.}~\bibnamefont {Kohri}},\ }\href
  {\doibase 10.1093/mnras/stae1098} {\bibfield  {journal} {\bibinfo  {journal}
  {Mon. Not. Roy. Astron. Soc.}\ }\textbf {\bibinfo {volume} {532}},\ \bibinfo
  {pages} {451} (\bibinfo {year} {2024})},\ \Eprint
  {http://arxiv.org/abs/2402.17823} {arXiv:2402.17823 [astro-ph.CO]}
  \BibitemShut {NoStop}%
\bibitem [{\citenamefont {Haque}\ \emph {et~al.}(2024)\citenamefont {Haque},
  \citenamefont {Maity}, \citenamefont {Maity},\ and\ \citenamefont
  {Mambrini}}]{Haque:2024eyh}%
  \BibitemOpen
  \bibfield  {author} {\bibinfo {author} {\bibfnamefont {M.~R.}\ \bibnamefont
  {Haque}}, \bibinfo {author} {\bibfnamefont {S.}~\bibnamefont {Maity}},
  \bibinfo {author} {\bibfnamefont {D.}~\bibnamefont {Maity}}, \ and\ \bibinfo
  {author} {\bibfnamefont {Y.}~\bibnamefont {Mambrini}},\ }\href {\doibase
  10.1088/1475-7516/2024/07/002} {\bibfield  {journal} {\bibinfo  {journal}
  {JCAP}\ }\textbf {\bibinfo {volume} {07}},\ \bibinfo {pages} {002} (\bibinfo
  {year} {2024})},\ \Eprint {http://arxiv.org/abs/2404.16815} {arXiv:2404.16815
  [hep-ph]} \BibitemShut {NoStop}%
\bibitem [{\citenamefont {Chianese}\ \emph {et~al.}(2025)\citenamefont
  {Chianese}, \citenamefont {Boccia}, \citenamefont {Iocco}, \citenamefont
  {Miele},\ and\ \citenamefont {Saviano}}]{Chianese:2024rsn}%
  \BibitemOpen
  \bibfield  {author} {\bibinfo {author} {\bibfnamefont {M.}~\bibnamefont
  {Chianese}}, \bibinfo {author} {\bibfnamefont {A.}~\bibnamefont {Boccia}},
  \bibinfo {author} {\bibfnamefont {F.}~\bibnamefont {Iocco}}, \bibinfo
  {author} {\bibfnamefont {G.}~\bibnamefont {Miele}}, \ and\ \bibinfo {author}
  {\bibfnamefont {N.}~\bibnamefont {Saviano}},\ }\href {\doibase
  10.1103/PhysRevD.111.063036} {\bibfield  {journal} {\bibinfo  {journal}
  {Phys. Rev. D}\ }\textbf {\bibinfo {volume} {111}},\ \bibinfo {pages}
  {063036} (\bibinfo {year} {2025})},\ \Eprint
  {http://arxiv.org/abs/2410.07604} {arXiv:2410.07604 [astro-ph.HE]}
  \BibitemShut {NoStop}%
\bibitem [{\citenamefont {Chianese}(2025)}]{Chianese:2025wrk}%
  \BibitemOpen
  \bibfield  {author} {\bibinfo {author} {\bibfnamefont {M.}~\bibnamefont
  {Chianese}},\ }\href {\doibase 10.1103/cgbp-2hdy} {\bibfield  {journal}
  {\bibinfo  {journal} {Phys. Rev. D}\ }\textbf {\bibinfo {volume} {112}},\
  \bibinfo {pages} {023043} (\bibinfo {year} {2025})},\ \Eprint
  {http://arxiv.org/abs/2504.03838} {arXiv:2504.03838 [astro-ph.HE]}
  \BibitemShut {NoStop}%
\bibitem [{\citenamefont {Tan}\ and\ \citenamefont {Zhou}(2026)}]{Tan:2025vxp}%
  \BibitemOpen
  \bibfield  {author} {\bibinfo {author} {\bibfnamefont {X.-h.}\ \bibnamefont
  {Tan}}\ and\ \bibinfo {author} {\bibfnamefont {Y.-f.}\ \bibnamefont {Zhou}},\
  }\href {\doibase 10.1016/j.physletb.2026.140404} {\bibfield  {journal}
  {\bibinfo  {journal} {Phys. Lett. B}\ }\textbf {\bibinfo {volume} {876}},\
  \bibinfo {pages} {140404} (\bibinfo {year} {2026})},\ \Eprint
  {http://arxiv.org/abs/2505.19857} {arXiv:2505.19857 [astro-ph.CO]}
  \BibitemShut {NoStop}%
\bibitem [{\citenamefont {Dondarini}\ \emph {et~al.}(2025)\citenamefont
  {Dondarini}, \citenamefont {Marino}, \citenamefont {Panci},\ and\
  \citenamefont {Zantedeschi}}]{Dondarini:2025ktz}%
  \BibitemOpen
  \bibfield  {author} {\bibinfo {author} {\bibfnamefont {A.}~\bibnamefont
  {Dondarini}}, \bibinfo {author} {\bibfnamefont {G.}~\bibnamefont {Marino}},
  \bibinfo {author} {\bibfnamefont {P.}~\bibnamefont {Panci}}, \ and\ \bibinfo
  {author} {\bibfnamefont {M.}~\bibnamefont {Zantedeschi}},\ }\href {\doibase
  10.1088/1475-7516/2025/11/006} {\bibfield  {journal} {\bibinfo  {journal}
  {JCAP}\ }\textbf {\bibinfo {volume} {11}},\ \bibinfo {pages} {006} (\bibinfo
  {year} {2025})},\ \Eprint {http://arxiv.org/abs/2506.13861} {arXiv:2506.13861
  [hep-ph]} \BibitemShut {NoStop}%
\bibitem [{\citenamefont {Tseng}\ and\ \citenamefont
  {Yeh}(2026)}]{Tseng:2025fjf}%
  \BibitemOpen
  \bibfield  {author} {\bibinfo {author} {\bibfnamefont {P.-Y.}\ \bibnamefont
  {Tseng}}\ and\ \bibinfo {author} {\bibfnamefont {Y.-M.}\ \bibnamefont
  {Yeh}},\ }\href {\doibase 10.1140/epjc/s10052-026-15608-0} {\bibfield
  {journal} {\bibinfo  {journal} {Eur. Phys. J. C}\ }\textbf {\bibinfo {volume}
  {86}},\ \bibinfo {pages} {434} (\bibinfo {year} {2026})},\ \Eprint
  {http://arxiv.org/abs/2511.01848} {arXiv:2511.01848 [hep-ph]} \BibitemShut
  {NoStop}%
\bibitem [{\citenamefont {Chaudhuri}(2026)}]{Chaudhuri:2026ljw}%
  \BibitemOpen
  \bibfield  {author} {\bibinfo {author} {\bibfnamefont {A.}~\bibnamefont
  {Chaudhuri}},\ }\href@noop {} {\  (\bibinfo {year} {2026})},\ \Eprint
  {http://arxiv.org/abs/2604.06858} {arXiv:2604.06858 [hep-ph]} \BibitemShut
  {NoStop}%
\bibitem [{\citenamefont {Tan}\ \emph {et~al.}(2026)\citenamefont {Tan},
  \citenamefont {Xia},\ and\ \citenamefont {Zhou}}]{Tan:2026ybz}%
  \BibitemOpen
  \bibfield  {author} {\bibinfo {author} {\bibfnamefont {X.-H.}\ \bibnamefont
  {Tan}}, \bibinfo {author} {\bibfnamefont {J.-q.}\ \bibnamefont {Xia}}, \ and\
  \bibinfo {author} {\bibfnamefont {Y.-F.}\ \bibnamefont {Zhou}},\ }\href@noop
  {} {\  (\bibinfo {year} {2026})},\ \Eprint {http://arxiv.org/abs/2606.16145}
  {arXiv:2606.16145 [astro-ph.CO]} \BibitemShut {NoStop}%
\bibitem [{\citenamefont {Asl}\ \emph {et~al.}(2026)\citenamefont {Asl},
  \citenamefont {Choubey},\ and\ \citenamefont {Lund}}]{Asl:2026rcm}%
  \BibitemOpen
  \bibfield  {author} {\bibinfo {author} {\bibfnamefont {A.~M.}\ \bibnamefont
  {Asl}}, \bibinfo {author} {\bibfnamefont {S.}~\bibnamefont {Choubey}}, \ and\
  \bibinfo {author} {\bibfnamefont {A.}~\bibnamefont {Lund}},\ }\href@noop {}
  {\  (\bibinfo {year} {2026})},\ \Eprint {http://arxiv.org/abs/2608.08144}
  {arXiv:2608.08144 [hep-ph]} \BibitemShut {NoStop}%
\bibitem [{\citenamefont {Calabrese}\ \emph
  {et~al.}(2022{\natexlab{a}})\citenamefont {Calabrese}, \citenamefont
  {Chianese}, \citenamefont {Fiorillo},\ and\ \citenamefont
  {Saviano}}]{Calabrese:2021src}%
  \BibitemOpen
  \bibfield  {author} {\bibinfo {author} {\bibfnamefont {R.}~\bibnamefont
  {Calabrese}}, \bibinfo {author} {\bibfnamefont {M.}~\bibnamefont {Chianese}},
  \bibinfo {author} {\bibfnamefont {D.~F.~G.}\ \bibnamefont {Fiorillo}}, \ and\
  \bibinfo {author} {\bibfnamefont {N.}~\bibnamefont {Saviano}},\ }\href
  {\doibase 10.1103/PhysRevD.105.L021302} {\bibfield  {journal} {\bibinfo
  {journal} {Phys. Rev. D}\ }\textbf {\bibinfo {volume} {105}},\ \bibinfo
  {pages} {L021302} (\bibinfo {year} {2022}{\natexlab{a}})},\ \Eprint
  {http://arxiv.org/abs/2107.13001} {arXiv:2107.13001 [hep-ph]} \BibitemShut
  {NoStop}%
\bibitem [{\citenamefont {Li}\ and\ \citenamefont {Liao}(2022)}]{Li:2022jxo}%
  \BibitemOpen
  \bibfield  {author} {\bibinfo {author} {\bibfnamefont {T.}~\bibnamefont
  {Li}}\ and\ \bibinfo {author} {\bibfnamefont {J.}~\bibnamefont {Liao}},\
  }\href {\doibase 10.1103/PhysRevD.106.055043} {\bibfield  {journal} {\bibinfo
   {journal} {Phys. Rev. D}\ }\textbf {\bibinfo {volume} {106}},\ \bibinfo
  {pages} {055043} (\bibinfo {year} {2022})},\ \Eprint
  {http://arxiv.org/abs/2203.14443} {arXiv:2203.14443 [hep-ph]} \BibitemShut
  {NoStop}%
\bibitem [{\citenamefont {Calabrese}\ \emph
  {et~al.}(2022{\natexlab{b}})\citenamefont {Calabrese}, \citenamefont
  {Chianese}, \citenamefont {Fiorillo},\ and\ \citenamefont
  {Saviano}}]{Calabrese:2022rfa}%
  \BibitemOpen
  \bibfield  {author} {\bibinfo {author} {\bibfnamefont {R.}~\bibnamefont
  {Calabrese}}, \bibinfo {author} {\bibfnamefont {M.}~\bibnamefont {Chianese}},
  \bibinfo {author} {\bibfnamefont {D.~F.~G.}\ \bibnamefont {Fiorillo}}, \ and\
  \bibinfo {author} {\bibfnamefont {N.}~\bibnamefont {Saviano}},\ }\href
  {\doibase 10.1103/PhysRevD.105.103024} {\bibfield  {journal} {\bibinfo
  {journal} {Phys. Rev. D}\ }\textbf {\bibinfo {volume} {105}},\ \bibinfo
  {pages} {103024} (\bibinfo {year} {2022}{\natexlab{b}})},\ \Eprint
  {http://arxiv.org/abs/2203.17093} {arXiv:2203.17093 [hep-ph]} \BibitemShut
  {NoStop}%
\bibitem [{\citenamefont {Zhang}\ \emph {et~al.}(2023)\citenamefont {Zhang}
  \emph {et~al.}}]{CDEX:2022dda}%
  \BibitemOpen
  \bibfield  {author} {\bibinfo {author} {\bibfnamefont {Z.~H.}\ \bibnamefont
  {Zhang}} \emph {et~al.} (\bibinfo {collaboration} {CDEX}),\ }\href {\doibase
  10.1103/PhysRevD.108.052006} {\bibfield  {journal} {\bibinfo  {journal}
  {Phys. Rev. D}\ }\textbf {\bibinfo {volume} {108}},\ \bibinfo {pages}
  {052006} (\bibinfo {year} {2023})},\ \Eprint
  {http://arxiv.org/abs/2211.07477} {arXiv:2211.07477 [hep-ex]} \BibitemShut
  {NoStop}%
\bibitem [{\citenamefont {Li}\ and\ \citenamefont {Zhang}(2022)}]{Li:2022xqh}%
  \BibitemOpen
  \bibfield  {author} {\bibinfo {author} {\bibfnamefont {T.}~\bibnamefont
  {Li}}\ and\ \bibinfo {author} {\bibfnamefont {R.-J.}\ \bibnamefont {Zhang}},\
  }\href {\doibase 10.1103/PhysRevD.106.095034} {\bibfield  {journal} {\bibinfo
   {journal} {Phys. Rev. D}\ }\textbf {\bibinfo {volume} {106}},\ \bibinfo
  {pages} {095034} (\bibinfo {year} {2022})},\ \Eprint
  {http://arxiv.org/abs/2208.02696} {arXiv:2208.02696 [hep-ph]} \BibitemShut
  {NoStop}%
\bibitem [{\citenamefont {Jho}\ \emph {et~al.}(2026)\citenamefont {Jho},
  \citenamefont {Kim}, \citenamefont {Park}, \citenamefont {Park},\ and\
  \citenamefont {Park}}]{Jho:2022wxd}%
  \BibitemOpen
  \bibfield  {author} {\bibinfo {author} {\bibfnamefont {Y.}~\bibnamefont
  {Jho}}, \bibinfo {author} {\bibfnamefont {T.-G.}\ \bibnamefont {Kim}},
  \bibinfo {author} {\bibfnamefont {J.-C.}\ \bibnamefont {Park}}, \bibinfo
  {author} {\bibfnamefont {S.~C.}\ \bibnamefont {Park}}, \ and\ \bibinfo
  {author} {\bibfnamefont {Y.}~\bibnamefont {Park}},\ }\href {\doibase
  10.1093/ptep/ptag011} {\bibfield  {journal} {\bibinfo  {journal} {PTEP}\
  }\textbf {\bibinfo {volume} {2026}},\ \bibinfo {pages} {023B07} (\bibinfo
  {year} {2026})},\ \Eprint {http://arxiv.org/abs/2212.11977} {arXiv:2212.11977
  [hep-ph]} \BibitemShut {NoStop}%
\bibitem [{\citenamefont {Agashe}\ \emph {et~al.}(2023)\citenamefont {Agashe},
  \citenamefont {Chang}, \citenamefont {Clark}, \citenamefont {Dutta},
  \citenamefont {Tsai},\ and\ \citenamefont {Xu}}]{Agashe:2022phd}%
  \BibitemOpen
  \bibfield  {author} {\bibinfo {author} {\bibfnamefont {K.}~\bibnamefont
  {Agashe}}, \bibinfo {author} {\bibfnamefont {J.~H.}\ \bibnamefont {Chang}},
  \bibinfo {author} {\bibfnamefont {S.~J.}\ \bibnamefont {Clark}}, \bibinfo
  {author} {\bibfnamefont {B.}~\bibnamefont {Dutta}}, \bibinfo {author}
  {\bibfnamefont {Y.}~\bibnamefont {Tsai}}, \ and\ \bibinfo {author}
  {\bibfnamefont {T.}~\bibnamefont {Xu}},\ }\href {\doibase
  10.1103/PhysRevD.108.023014} {\bibfield  {journal} {\bibinfo  {journal}
  {Phys. Rev. D}\ }\textbf {\bibinfo {volume} {108}},\ \bibinfo {pages}
  {023014} (\bibinfo {year} {2023})},\ \Eprint
  {http://arxiv.org/abs/2212.11980} {arXiv:2212.11980 [hep-ph]} \BibitemShut
  {NoStop}%
\bibitem [{\citenamefont {De~Romeri}\ \emph {et~al.}(2025)\citenamefont
  {De~Romeri}, \citenamefont {Perez-Gonzalez},\ and\ \citenamefont
  {Tolino}}]{DeRomeri:2024zqs}%
  \BibitemOpen
  \bibfield  {author} {\bibinfo {author} {\bibfnamefont {V.}~\bibnamefont
  {De~Romeri}}, \bibinfo {author} {\bibfnamefont {Y.~F.}\ \bibnamefont
  {Perez-Gonzalez}}, \ and\ \bibinfo {author} {\bibfnamefont {A.}~\bibnamefont
  {Tolino}},\ }\href {\doibase 10.1088/1475-7516/2025/04/018} {\bibfield
  {journal} {\bibinfo  {journal} {JCAP}\ }\textbf {\bibinfo {volume} {04}},\
  \bibinfo {pages} {018} (\bibinfo {year} {2025})},\ \Eprint
  {http://arxiv.org/abs/2405.00124} {arXiv:2405.00124 [hep-ph]} \BibitemShut
  {NoStop}%
\bibitem [{\citenamefont {Federico}\ and\ \citenamefont
  {Profumo}(2025)}]{Federico:2024fyt}%
  \BibitemOpen
  \bibfield  {author} {\bibinfo {author} {\bibfnamefont {K.}~\bibnamefont
  {Federico}}\ and\ \bibinfo {author} {\bibfnamefont {S.}~\bibnamefont
  {Profumo}},\ }\href {\doibase 10.1103/PhysRevD.111.063006} {\bibfield
  {journal} {\bibinfo  {journal} {Phys. Rev. D}\ }\textbf {\bibinfo {volume}
  {111}},\ \bibinfo {pages} {063006} (\bibinfo {year} {2025})},\ \Eprint
  {http://arxiv.org/abs/2411.17038} {arXiv:2411.17038 [hep-ph]} \BibitemShut
  {NoStop}%
\bibitem [{\citenamefont {Cheung}\ \emph {et~al.}(2025)\citenamefont {Cheung},
  \citenamefont {Ouseph}, \citenamefont {Tseng},\ and\ \citenamefont
  {Kang}}]{Cheung:2025gdn}%
  \BibitemOpen
  \bibfield  {author} {\bibinfo {author} {\bibfnamefont {K.}~\bibnamefont
  {Cheung}}, \bibinfo {author} {\bibfnamefont {C.~J.}\ \bibnamefont {Ouseph}},
  \bibinfo {author} {\bibfnamefont {P.-Y.}\ \bibnamefont {Tseng}}, \ and\
  \bibinfo {author} {\bibfnamefont {S.~K.}\ \bibnamefont {Kang}},\ }\href@noop
  {} {\  (\bibinfo {year} {2025})},\ \Eprint {http://arxiv.org/abs/2503.04175}
  {arXiv:2503.04175 [hep-ph]} \BibitemShut {NoStop}%
\bibitem [{\citenamefont {Li}\ and\ \citenamefont {Zhang}(2025)}]{Li:2025lip}%
  \BibitemOpen
  \bibfield  {author} {\bibinfo {author} {\bibfnamefont {T.}~\bibnamefont
  {Li}}\ and\ \bibinfo {author} {\bibfnamefont {R.-J.}\ \bibnamefont {Zhang}},\
  }\href {\doibase 10.1103/fxl6-mcxw} {\bibfield  {journal} {\bibinfo
  {journal} {Phys. Rev. D}\ }\textbf {\bibinfo {volume} {112}},\ \bibinfo
  {pages} {035028} (\bibinfo {year} {2025})},\ \Eprint
  {http://arxiv.org/abs/2504.14185} {arXiv:2504.14185 [hep-ph]} \BibitemShut
  {NoStop}%
\bibitem [{\citenamefont {Zhu}\ \emph {et~al.}(2026)\citenamefont {Zhu},
  \citenamefont {Jiang}, \citenamefont {Li},\ and\ \citenamefont
  {Liao}}]{Zhu:2026ush}%
  \BibitemOpen
  \bibfield  {author} {\bibinfo {author} {\bibfnamefont {T.}~\bibnamefont
  {Zhu}}, \bibinfo {author} {\bibfnamefont {C.-R.}\ \bibnamefont {Jiang}},
  \bibinfo {author} {\bibfnamefont {T.}~\bibnamefont {Li}}, \ and\ \bibinfo
  {author} {\bibfnamefont {J.}~\bibnamefont {Liao}},\ }\href {\doibase
  10.1103/m2ly-213l} {\bibfield  {journal} {\bibinfo  {journal} {Phys. Rev. D}\
  }\textbf {\bibinfo {volume} {113}},\ \bibinfo {pages} {075018} (\bibinfo
  {year} {2026})},\ \Eprint {http://arxiv.org/abs/2602.03010} {arXiv:2602.03010
  [hep-ph]} \BibitemShut {NoStop}%
\bibitem [{\citenamefont {Jeesun}\ \emph {et~al.}(2026)\citenamefont {Jeesun},
  \citenamefont {Majumdar},\ and\ \citenamefont {Srivastava}}]{Jeesun:2026wcx}%
  \BibitemOpen
  \bibfield  {author} {\bibinfo {author} {\bibfnamefont {S.}~\bibnamefont
  {Jeesun}}, \bibinfo {author} {\bibfnamefont {A.}~\bibnamefont {Majumdar}}, \
  and\ \bibinfo {author} {\bibfnamefont {R.}~\bibnamefont {Srivastava}},\
  }\href {\doibase 10.1088/1475-7516/2026/06/063} {\bibfield  {journal}
  {\bibinfo  {journal} {JCAP}\ }\textbf {\bibinfo {volume} {06}},\ \bibinfo
  {pages} {063} (\bibinfo {year} {2026})},\ \Eprint
  {http://arxiv.org/abs/2602.04858} {arXiv:2602.04858 [hep-ph]} \BibitemShut
  {NoStop}%
\bibitem [{\citenamefont {Peccei}\ and\ \citenamefont
  {Quinn}(1977{\natexlab{a}})}]{Peccei:1977hh}%
  \BibitemOpen
  \bibfield  {author} {\bibinfo {author} {\bibfnamefont {R.~D.}\ \bibnamefont
  {Peccei}}\ and\ \bibinfo {author} {\bibfnamefont {H.~R.}\ \bibnamefont
  {Quinn}},\ }\href {\doibase 10.1103/PhysRevLett.38.1440} {\bibfield
  {journal} {\bibinfo  {journal} {Phys. Rev. Lett.}\ }\textbf {\bibinfo
  {volume} {38}},\ \bibinfo {pages} {1440} (\bibinfo {year}
  {1977}{\natexlab{a}})}\BibitemShut {NoStop}%
\bibitem [{\citenamefont {Peccei}\ and\ \citenamefont
  {Quinn}(1977{\natexlab{b}})}]{Peccei:1977ur}%
  \BibitemOpen
  \bibfield  {author} {\bibinfo {author} {\bibfnamefont {R.~D.}\ \bibnamefont
  {Peccei}}\ and\ \bibinfo {author} {\bibfnamefont {H.~R.}\ \bibnamefont
  {Quinn}},\ }\href {\doibase 10.1103/PhysRevD.16.1791} {\bibfield  {journal}
  {\bibinfo  {journal} {Phys. Rev. D}\ }\textbf {\bibinfo {volume} {16}},\
  \bibinfo {pages} {1791} (\bibinfo {year} {1977}{\natexlab{b}})}\BibitemShut
  {NoStop}%
\bibitem [{\citenamefont {Weinberg}(1978)}]{Weinberg:1977ma}%
  \BibitemOpen
  \bibfield  {author} {\bibinfo {author} {\bibfnamefont {S.}~\bibnamefont
  {Weinberg}},\ }\href {\doibase 10.1103/PhysRevLett.40.223} {\bibfield
  {journal} {\bibinfo  {journal} {Phys. Rev. Lett.}\ }\textbf {\bibinfo
  {volume} {40}},\ \bibinfo {pages} {223} (\bibinfo {year} {1978})}\BibitemShut
  {NoStop}%
\bibitem [{\citenamefont {Wilczek}(1978)}]{Wilczek:1977pj}%
  \BibitemOpen
  \bibfield  {author} {\bibinfo {author} {\bibfnamefont {F.}~\bibnamefont
  {Wilczek}},\ }\href {\doibase 10.1103/PhysRevLett.40.279} {\bibfield
  {journal} {\bibinfo  {journal} {Phys. Rev. Lett.}\ }\textbf {\bibinfo
  {volume} {40}},\ \bibinfo {pages} {279} (\bibinfo {year} {1978})}\BibitemShut
  {NoStop}%
\bibitem [{\citenamefont {Dimopoulos}(1979)}]{Dimopoulos:1979pp}%
  \BibitemOpen
  \bibfield  {author} {\bibinfo {author} {\bibfnamefont {S.}~\bibnamefont
  {Dimopoulos}},\ }\href {\doibase 10.1016/0370-2693(79)91233-4} {\bibfield
  {journal} {\bibinfo  {journal} {Phys. Lett. B}\ }\textbf {\bibinfo {volume}
  {84}},\ \bibinfo {pages} {435} (\bibinfo {year} {1979})}\BibitemShut
  {NoStop}%
\bibitem [{\citenamefont {Tye}(1981)}]{Tye:1981zy}%
  \BibitemOpen
  \bibfield  {author} {\bibinfo {author} {\bibfnamefont {S.~H.~H.}\
  \bibnamefont {Tye}},\ }\href {\doibase 10.1103/PhysRevLett.47.1035}
  {\bibfield  {journal} {\bibinfo  {journal} {Phys. Rev. Lett.}\ }\textbf
  {\bibinfo {volume} {47}},\ \bibinfo {pages} {1035} (\bibinfo {year}
  {1981})}\BibitemShut {NoStop}%
\bibitem [{\citenamefont {Zhitnitsky}(1980)}]{Zhitnitsky:1980tq}%
  \BibitemOpen
  \bibfield  {author} {\bibinfo {author} {\bibfnamefont {A.~R.}\ \bibnamefont
  {Zhitnitsky}},\ }\href@noop {} {\bibfield  {journal} {\bibinfo  {journal}
  {Sov. J. Nucl. Phys.}\ }\textbf {\bibinfo {volume} {31}},\ \bibinfo {pages}
  {260} (\bibinfo {year} {1980})}\BibitemShut {NoStop}%
\bibitem [{\citenamefont {Dine}\ \emph {et~al.}(1981)\citenamefont {Dine},
  \citenamefont {Fischler},\ and\ \citenamefont {Srednicki}}]{Dine:1981rt}%
  \BibitemOpen
  \bibfield  {author} {\bibinfo {author} {\bibfnamefont {M.}~\bibnamefont
  {Dine}}, \bibinfo {author} {\bibfnamefont {W.}~\bibnamefont {Fischler}}, \
  and\ \bibinfo {author} {\bibfnamefont {M.}~\bibnamefont {Srednicki}},\ }\href
  {\doibase 10.1016/0370-2693(81)90590-6} {\bibfield  {journal} {\bibinfo
  {journal} {Phys. Lett. B}\ }\textbf {\bibinfo {volume} {104}},\ \bibinfo
  {pages} {199} (\bibinfo {year} {1981})}\BibitemShut {NoStop}%
\bibitem [{\citenamefont {Holdom}\ and\ \citenamefont
  {Peskin}(1982)}]{Holdom:1982ex}%
  \BibitemOpen
  \bibfield  {author} {\bibinfo {author} {\bibfnamefont {B.}~\bibnamefont
  {Holdom}}\ and\ \bibinfo {author} {\bibfnamefont {M.~E.}\ \bibnamefont
  {Peskin}},\ }\href {\doibase 10.1016/0550-3213(82)90228-0} {\bibfield
  {journal} {\bibinfo  {journal} {Nucl. Phys. B}\ }\textbf {\bibinfo {volume}
  {208}},\ \bibinfo {pages} {397} (\bibinfo {year} {1982})}\BibitemShut
  {NoStop}%
\bibitem [{\citenamefont {Kaplan}(1985)}]{Kaplan:1985dv}%
  \BibitemOpen
  \bibfield  {author} {\bibinfo {author} {\bibfnamefont {D.~B.}\ \bibnamefont
  {Kaplan}},\ }\href {\doibase 10.1016/0550-3213(85)90319-0} {\bibfield
  {journal} {\bibinfo  {journal} {Nucl. Phys. B}\ }\textbf {\bibinfo {volume}
  {260}},\ \bibinfo {pages} {215} (\bibinfo {year} {1985})}\BibitemShut
  {NoStop}%
\bibitem [{\citenamefont {Srednicki}(1985)}]{Srednicki:1985xd}%
  \BibitemOpen
  \bibfield  {author} {\bibinfo {author} {\bibfnamefont {M.}~\bibnamefont
  {Srednicki}},\ }\href {\doibase 10.1016/0550-3213(85)90054-9} {\bibfield
  {journal} {\bibinfo  {journal} {Nucl. Phys. B}\ }\textbf {\bibinfo {volume}
  {260}},\ \bibinfo {pages} {689} (\bibinfo {year} {1985})}\BibitemShut
  {NoStop}%
\bibitem [{\citenamefont {Flynn}\ and\ \citenamefont
  {Randall}(1987)}]{Flynn:1987rs}%
  \BibitemOpen
  \bibfield  {author} {\bibinfo {author} {\bibfnamefont {J.~M.}\ \bibnamefont
  {Flynn}}\ and\ \bibinfo {author} {\bibfnamefont {L.}~\bibnamefont
  {Randall}},\ }\href {\doibase 10.1016/0550-3213(87)90089-7} {\bibfield
  {journal} {\bibinfo  {journal} {Nucl. Phys. B}\ }\textbf {\bibinfo {volume}
  {293}},\ \bibinfo {pages} {731} (\bibinfo {year} {1987})}\BibitemShut
  {NoStop}%
\bibitem [{\citenamefont {Kamionkowski}\ and\ \citenamefont
  {March-Russell}(1992)}]{Kamionkowski:1992mf}%
  \BibitemOpen
  \bibfield  {author} {\bibinfo {author} {\bibfnamefont {M.}~\bibnamefont
  {Kamionkowski}}\ and\ \bibinfo {author} {\bibfnamefont {J.}~\bibnamefont
  {March-Russell}},\ }\href {\doibase 10.1016/0370-2693(92)90492-M} {\bibfield
  {journal} {\bibinfo  {journal} {Phys. Lett. B}\ }\textbf {\bibinfo {volume}
  {282}},\ \bibinfo {pages} {137} (\bibinfo {year} {1992})},\ \Eprint
  {http://arxiv.org/abs/hep-th/9202003} {arXiv:hep-th/9202003} \BibitemShut
  {NoStop}%
\bibitem [{\citenamefont {Berezhiani}\ \emph {et~al.}(2001)\citenamefont
  {Berezhiani}, \citenamefont {Gianfagna},\ and\ \citenamefont
  {Giannotti}}]{Berezhiani:2000gh}%
  \BibitemOpen
  \bibfield  {author} {\bibinfo {author} {\bibfnamefont {Z.}~\bibnamefont
  {Berezhiani}}, \bibinfo {author} {\bibfnamefont {L.}~\bibnamefont
  {Gianfagna}}, \ and\ \bibinfo {author} {\bibfnamefont {M.}~\bibnamefont
  {Giannotti}},\ }\href {\doibase 10.1016/S0370-2693(00)01392-7} {\bibfield
  {journal} {\bibinfo  {journal} {Phys. Lett. B}\ }\textbf {\bibinfo {volume}
  {500}},\ \bibinfo {pages} {286} (\bibinfo {year} {2001})},\ \Eprint
  {http://arxiv.org/abs/hep-ph/0009290} {arXiv:hep-ph/0009290} \BibitemShut
  {NoStop}%
\bibitem [{\citenamefont {Hsu}\ and\ \citenamefont
  {Sannino}(2005)}]{Hsu:2004mf}%
  \BibitemOpen
  \bibfield  {author} {\bibinfo {author} {\bibfnamefont {S.~D.~H.}\
  \bibnamefont {Hsu}}\ and\ \bibinfo {author} {\bibfnamefont {F.}~\bibnamefont
  {Sannino}},\ }\href {\doibase 10.1016/j.physletb.2004.11.040} {\bibfield
  {journal} {\bibinfo  {journal} {Phys. Lett. B}\ }\textbf {\bibinfo {volume}
  {605}},\ \bibinfo {pages} {369} (\bibinfo {year} {2005})},\ \Eprint
  {http://arxiv.org/abs/hep-ph/0408319} {arXiv:hep-ph/0408319} \BibitemShut
  {NoStop}%
\bibitem [{\citenamefont {Hook}(2015)}]{Hook:2014cda}%
  \BibitemOpen
  \bibfield  {author} {\bibinfo {author} {\bibfnamefont {A.}~\bibnamefont
  {Hook}},\ }\href {\doibase 10.1103/PhysRevLett.114.141801} {\bibfield
  {journal} {\bibinfo  {journal} {Phys. Rev. Lett.}\ }\textbf {\bibinfo
  {volume} {114}},\ \bibinfo {pages} {141801} (\bibinfo {year} {2015})},\
  \Eprint {http://arxiv.org/abs/1411.3325} {arXiv:1411.3325 [hep-ph]}
  \BibitemShut {NoStop}%
\bibitem [{\citenamefont {Alonso-{\'A}lvarez}\ \emph
  {et~al.}(2019)\citenamefont {Alonso-{\'A}lvarez}, \citenamefont {Gavela},\
  and\ \citenamefont {Quilez}}]{Alonso-Alvarez:2018irt}%
  \BibitemOpen
  \bibfield  {author} {\bibinfo {author} {\bibfnamefont {G.}~\bibnamefont
  {Alonso-{\'A}lvarez}}, \bibinfo {author} {\bibfnamefont {M.~B.}\ \bibnamefont
  {Gavela}}, \ and\ \bibinfo {author} {\bibfnamefont {P.}~\bibnamefont
  {Quilez}},\ }\href {\doibase 10.1140/epjc/s10052-019-6732-5} {\bibfield
  {journal} {\bibinfo  {journal} {Eur. Phys. J. C}\ }\textbf {\bibinfo {volume}
  {79}},\ \bibinfo {pages} {223} (\bibinfo {year} {2019})},\ \Eprint
  {http://arxiv.org/abs/1811.05466} {arXiv:1811.05466 [hep-ph]} \BibitemShut
  {NoStop}%
\bibitem [{\citenamefont {Hook}\ \emph {et~al.}(2020)\citenamefont {Hook},
  \citenamefont {Kumar}, \citenamefont {Liu},\ and\ \citenamefont
  {Sundrum}}]{Hook:2019qoh}%
  \BibitemOpen
  \bibfield  {author} {\bibinfo {author} {\bibfnamefont {A.}~\bibnamefont
  {Hook}}, \bibinfo {author} {\bibfnamefont {S.}~\bibnamefont {Kumar}},
  \bibinfo {author} {\bibfnamefont {Z.}~\bibnamefont {Liu}}, \ and\ \bibinfo
  {author} {\bibfnamefont {R.}~\bibnamefont {Sundrum}},\ }\href {\doibase
  10.1103/PhysRevLett.124.221801} {\bibfield  {journal} {\bibinfo  {journal}
  {Phys. Rev. Lett.}\ }\textbf {\bibinfo {volume} {124}},\ \bibinfo {pages}
  {221801} (\bibinfo {year} {2020})},\ \Eprint
  {http://arxiv.org/abs/1911.12364} {arXiv:1911.12364 [hep-ph]} \BibitemShut
  {NoStop}%
\bibitem [{\citenamefont {Caputo}\ \emph {et~al.}(2019)\citenamefont {Caputo}
  \emph {et~al.}}]{AMEGO:2019gny}%
  \BibitemOpen
  \bibfield  {author} {\bibinfo {author} {\bibfnamefont {R.}~\bibnamefont
  {Caputo}} \emph {et~al.} (\bibinfo {collaboration} {AMEGO}),\ }\href@noop {}
  {\  (\bibinfo {year} {2019})},\ \Eprint {http://arxiv.org/abs/1907.07558}
  {arXiv:1907.07558 [astro-ph.IM]} \BibitemShut {NoStop}%
\bibitem [{\citenamefont {Kierans}(2020)}]{Kierans:2020otl}%
  \BibitemOpen
  \bibfield  {author} {\bibinfo {author} {\bibfnamefont {C.~A.}\ \bibnamefont
  {Kierans}} (\bibinfo {collaboration} {AMEGO Team}),\ }\href {\doibase
  10.1117/12.2562352} {\bibfield  {journal} {\bibinfo  {journal} {Proc. SPIE
  Int. Soc. Opt. Eng.}\ }\textbf {\bibinfo {volume} {11444}},\ \bibinfo {pages}
  {1144431} (\bibinfo {year} {2020})},\ \Eprint
  {http://arxiv.org/abs/2101.03105} {arXiv:2101.03105 [astro-ph.IM]}
  \BibitemShut {NoStop}%
\bibitem [{\citenamefont {De~Angelis}\ \emph {et~al.}(2017)\citenamefont
  {De~Angelis} \emph {et~al.}}]{e-ASTROGAM:2016bph}%
  \BibitemOpen
  \bibfield  {author} {\bibinfo {author} {\bibfnamefont {A.}~\bibnamefont
  {De~Angelis}} \emph {et~al.} (\bibinfo {collaboration} {e-ASTROGAM}),\ }\href
  {\doibase 10.1007/s10686-017-9533-6} {\bibfield  {journal} {\bibinfo
  {journal} {Exper. Astron.}\ }\textbf {\bibinfo {volume} {44}},\ \bibinfo
  {pages} {25} (\bibinfo {year} {2017})},\ \Eprint
  {http://arxiv.org/abs/1611.02232} {arXiv:1611.02232 [astro-ph.HE]}
  \BibitemShut {NoStop}%
\bibitem [{\citenamefont {Tavani}\ \emph {et~al.}(2018)\citenamefont {Tavani}
  \emph {et~al.}}]{e-ASTROGAM:2017pxr}%
  \BibitemOpen
  \bibfield  {author} {\bibinfo {author} {\bibfnamefont {M.}~\bibnamefont
  {Tavani}} \emph {et~al.} (\bibinfo {collaboration} {e-ASTROGAM}),\ }\href
  {\doibase 10.1016/j.jheap.2018.07.001} {\bibfield  {journal} {\bibinfo
  {journal} {JHEAp}\ }\textbf {\bibinfo {volume} {19}},\ \bibinfo {pages} {1}
  (\bibinfo {year} {2018})},\ \Eprint {http://arxiv.org/abs/1711.01265}
  {arXiv:1711.01265 [astro-ph.HE]} \BibitemShut {NoStop}%
\bibitem [{\citenamefont {Dzhatdoev}\ and\ \citenamefont
  {Podlesnyi}(2019)}]{Dzhatdoev:2019kay}%
  \BibitemOpen
  \bibfield  {author} {\bibinfo {author} {\bibfnamefont {T.}~\bibnamefont
  {Dzhatdoev}}\ and\ \bibinfo {author} {\bibfnamefont {E.}~\bibnamefont
  {Podlesnyi}},\ }\href {\doibase 10.1016/j.astropartphys.2019.04.004}
  {\bibfield  {journal} {\bibinfo  {journal} {Astropart. Phys.}\ }\textbf
  {\bibinfo {volume} {112}},\ \bibinfo {pages} {1} (\bibinfo {year} {2019})},\
  \Eprint {http://arxiv.org/abs/1902.01491} {arXiv:1902.01491 [astro-ph.HE]}
  \BibitemShut {NoStop}%
\bibitem [{\citenamefont {Page}(1976{\natexlab{a}})}]{Page:1976df}%
  \BibitemOpen
  \bibfield  {author} {\bibinfo {author} {\bibfnamefont {D.~N.}\ \bibnamefont
  {Page}},\ }\href {\doibase 10.1103/PhysRevD.13.198} {\bibfield  {journal}
  {\bibinfo  {journal} {Phys. Rev. D}\ }\textbf {\bibinfo {volume} {13}},\
  \bibinfo {pages} {198} (\bibinfo {year} {1976}{\natexlab{a}})}\BibitemShut
  {NoStop}%
\bibitem [{\citenamefont {Page}(1977)}]{Page:1977um}%
  \BibitemOpen
  \bibfield  {author} {\bibinfo {author} {\bibfnamefont {D.~N.}\ \bibnamefont
  {Page}},\ }\href {\doibase 10.1103/PhysRevD.16.2402} {\bibfield  {journal}
  {\bibinfo  {journal} {Phys. Rev. D}\ }\textbf {\bibinfo {volume} {16}},\
  \bibinfo {pages} {2402} (\bibinfo {year} {1977})}\BibitemShut {NoStop}%
\bibitem [{\citenamefont {MacGibbon}\ and\ \citenamefont
  {Webber}(1990)}]{MacGibbon:1990zk}%
  \BibitemOpen
  \bibfield  {author} {\bibinfo {author} {\bibfnamefont {J.~H.}\ \bibnamefont
  {MacGibbon}}\ and\ \bibinfo {author} {\bibfnamefont {B.~R.}\ \bibnamefont
  {Webber}},\ }\href {\doibase 10.1103/PhysRevD.41.3052} {\bibfield  {journal}
  {\bibinfo  {journal} {Phys. Rev. D}\ }\textbf {\bibinfo {volume} {41}},\
  \bibinfo {pages} {3052} (\bibinfo {year} {1990})}\BibitemShut {NoStop}%
\bibitem [{\citenamefont {MacGibbon}(1991)}]{MacGibbon:1991tj}%
  \BibitemOpen
  \bibfield  {author} {\bibinfo {author} {\bibfnamefont {J.~H.}\ \bibnamefont
  {MacGibbon}},\ }\href {\doibase 10.1103/PhysRevD.44.376} {\bibfield
  {journal} {\bibinfo  {journal} {Phys. Rev. D}\ }\textbf {\bibinfo {volume}
  {44}},\ \bibinfo {pages} {376} (\bibinfo {year} {1991})}\BibitemShut
  {NoStop}%
\bibitem [{\citenamefont {Hawking}(1971)}]{Hawking:1971ei}%
  \BibitemOpen
  \bibfield  {author} {\bibinfo {author} {\bibfnamefont {S.}~\bibnamefont
  {Hawking}},\ }\href {\doibase 10.1093/mnras/152.1.75} {\bibfield  {journal}
  {\bibinfo  {journal} {Mon. Not. Roy. Astron. Soc.}\ }\textbf {\bibinfo
  {volume} {152}},\ \bibinfo {pages} {75} (\bibinfo {year} {1971})}\BibitemShut
  {NoStop}%
\bibitem [{\citenamefont {Page}(1976{\natexlab{b}})}]{Page:1976ki}%
  \BibitemOpen
  \bibfield  {author} {\bibinfo {author} {\bibfnamefont {D.~N.}\ \bibnamefont
  {Page}},\ }\href {\doibase 10.1103/PhysRevD.14.3260} {\bibfield  {journal}
  {\bibinfo  {journal} {Phys. Rev. D}\ }\textbf {\bibinfo {volume} {14}},\
  \bibinfo {pages} {3260} (\bibinfo {year} {1976}{\natexlab{b}})}\BibitemShut
  {NoStop}%
\bibitem [{\citenamefont {Baldes}\ \emph {et~al.}(2020)\citenamefont {Baldes},
  \citenamefont {Decant}, \citenamefont {Hooper},\ and\ \citenamefont
  {Lopez-Honorez}}]{Baldes:2020nuv}%
  \BibitemOpen
  \bibfield  {author} {\bibinfo {author} {\bibfnamefont {I.}~\bibnamefont
  {Baldes}}, \bibinfo {author} {\bibfnamefont {Q.}~\bibnamefont {Decant}},
  \bibinfo {author} {\bibfnamefont {D.~C.}\ \bibnamefont {Hooper}}, \ and\
  \bibinfo {author} {\bibfnamefont {L.}~\bibnamefont {Lopez-Honorez}},\ }\href
  {\doibase 10.1088/1475-7516/2020/08/045} {\bibfield  {journal} {\bibinfo
  {journal} {JCAP}\ }\textbf {\bibinfo {volume} {08}},\ \bibinfo {pages} {045}
  (\bibinfo {year} {2020})},\ \Eprint {http://arxiv.org/abs/2004.14773}
  {arXiv:2004.14773 [astro-ph.CO]} \BibitemShut {NoStop}%
\bibitem [{\citenamefont {Cheek}\ \emph {et~al.}(2022)\citenamefont {Cheek},
  \citenamefont {Heurtier}, \citenamefont {Perez-Gonzalez},\ and\ \citenamefont
  {Turner}}]{Cheek:2021odj}%
  \BibitemOpen
  \bibfield  {author} {\bibinfo {author} {\bibfnamefont {A.}~\bibnamefont
  {Cheek}}, \bibinfo {author} {\bibfnamefont {L.}~\bibnamefont {Heurtier}},
  \bibinfo {author} {\bibfnamefont {Y.~F.}\ \bibnamefont {Perez-Gonzalez}}, \
  and\ \bibinfo {author} {\bibfnamefont {J.}~\bibnamefont {Turner}},\ }\href
  {\doibase 10.1103/PhysRevD.105.015022} {\bibfield  {journal} {\bibinfo
  {journal} {Phys. Rev. D}\ }\textbf {\bibinfo {volume} {105}},\ \bibinfo
  {pages} {015022} (\bibinfo {year} {2022})},\ \Eprint
  {http://arxiv.org/abs/2107.00013} {arXiv:2107.00013 [hep-ph]} \BibitemShut
  {NoStop}%
\bibitem [{\citenamefont {Arbey}\ and\ \citenamefont
  {Auffinger}(2019)}]{Arbey:2019mbc}%
  \BibitemOpen
  \bibfield  {author} {\bibinfo {author} {\bibfnamefont {A.}~\bibnamefont
  {Arbey}}\ and\ \bibinfo {author} {\bibfnamefont {J.}~\bibnamefont
  {Auffinger}},\ }\href {\doibase 10.1140/epjc/s10052-019-7161-1} {\bibfield
  {journal} {\bibinfo  {journal} {Eur. Phys. J. C}\ }\textbf {\bibinfo {volume}
  {79}},\ \bibinfo {pages} {693} (\bibinfo {year} {2019})},\ \Eprint
  {http://arxiv.org/abs/1905.04268} {arXiv:1905.04268 [gr-qc]} \BibitemShut
  {NoStop}%
\bibitem [{\citenamefont {Arbey}\ and\ \citenamefont
  {Auffinger}(2021)}]{Arbey:2021mbl}%
  \BibitemOpen
  \bibfield  {author} {\bibinfo {author} {\bibfnamefont {A.}~\bibnamefont
  {Arbey}}\ and\ \bibinfo {author} {\bibfnamefont {J.}~\bibnamefont
  {Auffinger}},\ }\href {\doibase 10.1140/epjc/s10052-021-09702-8} {\bibfield
  {journal} {\bibinfo  {journal} {Eur. Phys. J. C}\ }\textbf {\bibinfo {volume}
  {81}},\ \bibinfo {pages} {910} (\bibinfo {year} {2021})},\ \Eprint
  {http://arxiv.org/abs/2108.02737} {arXiv:2108.02737 [gr-qc]} \BibitemShut
  {NoStop}%
\bibitem [{\citenamefont {Ray}\ \emph {et~al.}(2021)\citenamefont {Ray},
  \citenamefont {Laha}, \citenamefont {Mu{\~n}oz},\ and\ \citenamefont
  {Caputo}}]{Ray:2021mxu}%
  \BibitemOpen
  \bibfield  {author} {\bibinfo {author} {\bibfnamefont {A.}~\bibnamefont
  {Ray}}, \bibinfo {author} {\bibfnamefont {R.}~\bibnamefont {Laha}}, \bibinfo
  {author} {\bibfnamefont {J.~B.}\ \bibnamefont {Mu{\~n}oz}}, \ and\ \bibinfo
  {author} {\bibfnamefont {R.}~\bibnamefont {Caputo}},\ }\href {\doibase
  10.1103/PhysRevD.104.023516} {\bibfield  {journal} {\bibinfo  {journal}
  {Phys. Rev. D}\ }\textbf {\bibinfo {volume} {104}},\ \bibinfo {pages}
  {023516} (\bibinfo {year} {2021})},\ \Eprint
  {http://arxiv.org/abs/2102.06714} {arXiv:2102.06714 [astro-ph.CO]}
  \BibitemShut {NoStop}%
\bibitem [{\citenamefont {Navarro}\ \emph {et~al.}(1997)\citenamefont
  {Navarro}, \citenamefont {Frenk},\ and\ \citenamefont
  {White}}]{Navarro:1996gj}%
  \BibitemOpen
  \bibfield  {author} {\bibinfo {author} {\bibfnamefont {J.~F.}\ \bibnamefont
  {Navarro}}, \bibinfo {author} {\bibfnamefont {C.~S.}\ \bibnamefont {Frenk}},
  \ and\ \bibinfo {author} {\bibfnamefont {S.~D.~M.}\ \bibnamefont {White}},\
  }\href {\doibase 10.1086/304888} {\bibfield  {journal} {\bibinfo  {journal}
  {Astrophys. J.}\ }\textbf {\bibinfo {volume} {490}},\ \bibinfo {pages} {493}
  (\bibinfo {year} {1997})},\ \Eprint {http://arxiv.org/abs/astro-ph/9611107}
  {arXiv:astro-ph/9611107} \BibitemShut {NoStop}%
\bibitem [{\citenamefont {Salucci}\ \emph {et~al.}(2010)\citenamefont
  {Salucci}, \citenamefont {Nesti}, \citenamefont {Gentile},\ and\
  \citenamefont {Martins}}]{Salucci:2010qr}%
  \BibitemOpen
  \bibfield  {author} {\bibinfo {author} {\bibfnamefont {P.}~\bibnamefont
  {Salucci}}, \bibinfo {author} {\bibfnamefont {F.}~\bibnamefont {Nesti}},
  \bibinfo {author} {\bibfnamefont {G.}~\bibnamefont {Gentile}}, \ and\
  \bibinfo {author} {\bibfnamefont {C.~F.}\ \bibnamefont {Martins}},\ }\href
  {\doibase 10.1051/0004-6361/201014385} {\bibfield  {journal} {\bibinfo
  {journal} {Astron. Astrophys.}\ }\textbf {\bibinfo {volume} {523}},\ \bibinfo
  {pages} {A83} (\bibinfo {year} {2010})},\ \Eprint
  {http://arxiv.org/abs/1003.3101} {arXiv:1003.3101 [astro-ph.GA]} \BibitemShut
  {NoStop}%
\bibitem [{\citenamefont {Yuan}\ \emph {et~al.}(2017)\citenamefont {Yuan},
  \citenamefont {Lin}, \citenamefont {Fang},\ and\ \citenamefont
  {Bi}}]{Yuan:2017ozr}%
  \BibitemOpen
  \bibfield  {author} {\bibinfo {author} {\bibfnamefont {Q.}~\bibnamefont
  {Yuan}}, \bibinfo {author} {\bibfnamefont {S.-J.}\ \bibnamefont {Lin}},
  \bibinfo {author} {\bibfnamefont {K.}~\bibnamefont {Fang}}, \ and\ \bibinfo
  {author} {\bibfnamefont {X.-J.}\ \bibnamefont {Bi}},\ }\href {\doibase
  10.1103/PhysRevD.95.083007} {\bibfield  {journal} {\bibinfo  {journal} {Phys.
  Rev. D}\ }\textbf {\bibinfo {volume} {95}},\ \bibinfo {pages} {083007}
  (\bibinfo {year} {2017})},\ \Eprint {http://arxiv.org/abs/1701.06149}
  {arXiv:1701.06149 [astro-ph.HE]} \BibitemShut {NoStop}%
\bibitem [{\citenamefont {Wang}\ \emph {et~al.}(2021)\citenamefont {Wang},
  \citenamefont {Xia}, \citenamefont {Zhang}, \citenamefont {Zhou},\ and\
  \citenamefont {Chang}}]{Wang:2020uvi}%
  \BibitemOpen
  \bibfield  {author} {\bibinfo {author} {\bibfnamefont {S.}~\bibnamefont
  {Wang}}, \bibinfo {author} {\bibfnamefont {D.-M.}\ \bibnamefont {Xia}},
  \bibinfo {author} {\bibfnamefont {X.}~\bibnamefont {Zhang}}, \bibinfo
  {author} {\bibfnamefont {S.}~\bibnamefont {Zhou}}, \ and\ \bibinfo {author}
  {\bibfnamefont {Z.}~\bibnamefont {Chang}},\ }\href {\doibase
  10.1103/PhysRevD.103.043010} {\bibfield  {journal} {\bibinfo  {journal}
  {Phys. Rev. D}\ }\textbf {\bibinfo {volume} {103}},\ \bibinfo {pages}
  {043010} (\bibinfo {year} {2021})},\ \Eprint
  {http://arxiv.org/abs/2010.16053} {arXiv:2010.16053 [hep-ph]} \BibitemShut
  {NoStop}%
\bibitem [{\citenamefont {Aghanim}\ \emph {et~al.}(2020)\citenamefont {Aghanim}
  \emph {et~al.}}]{Planck:2018vyg}%
  \BibitemOpen
  \bibfield  {author} {\bibinfo {author} {\bibfnamefont {N.}~\bibnamefont
  {Aghanim}} \emph {et~al.} (\bibinfo {collaboration} {Planck}),\ }\href
  {\doibase 10.1051/0004-6361/201833910} {\bibfield  {journal} {\bibinfo
  {journal} {Astron. Astrophys.}\ }\textbf {\bibinfo {volume} {641}},\ \bibinfo
  {pages} {A6} (\bibinfo {year} {2020})},\ \bibinfo {note} {[Erratum:
  Astron.Astrophys. 652, C4 (2021)]},\ \Eprint
  {http://arxiv.org/abs/1807.06209} {arXiv:1807.06209 [astro-ph.CO]}
  \BibitemShut {NoStop}%
\bibitem [{\citenamefont {Raffelt}\ and\ \citenamefont
  {Stodolsky}(1988)}]{Raffelt:1987im}%
  \BibitemOpen
  \bibfield  {author} {\bibinfo {author} {\bibfnamefont {G.}~\bibnamefont
  {Raffelt}}\ and\ \bibinfo {author} {\bibfnamefont {L.}~\bibnamefont
  {Stodolsky}},\ }\href {\doibase 10.1103/PhysRevD.37.1237} {\bibfield
  {journal} {\bibinfo  {journal} {Phys. Rev. D}\ }\textbf {\bibinfo {volume}
  {37}},\ \bibinfo {pages} {1237} (\bibinfo {year} {1988})}\BibitemShut
  {NoStop}%
\bibitem [{\citenamefont {Mirizzi}\ \emph {et~al.}(2008)\citenamefont
  {Mirizzi}, \citenamefont {Raffelt},\ and\ \citenamefont
  {Serpico}}]{Mirizzi:2006zy}%
  \BibitemOpen
  \bibfield  {author} {\bibinfo {author} {\bibfnamefont {A.}~\bibnamefont
  {Mirizzi}}, \bibinfo {author} {\bibfnamefont {G.~G.}\ \bibnamefont
  {Raffelt}}, \ and\ \bibinfo {author} {\bibfnamefont {P.~D.}\ \bibnamefont
  {Serpico}},\ }\href {\doibase 10.1007/978-3-540-73518-2_7} {\bibfield
  {journal} {\bibinfo  {journal} {Lect. Notes Phys.}\ }\textbf {\bibinfo
  {volume} {741}},\ \bibinfo {pages} {115} (\bibinfo {year} {2008})},\ \Eprint
  {http://arxiv.org/abs/astro-ph/0607415} {arXiv:astro-ph/0607415} \BibitemShut
  {NoStop}%
\bibitem [{\citenamefont {Galanti}\ and\ \citenamefont
  {Roncadelli}(2022)}]{Galanti:2022ijh}%
  \BibitemOpen
  \bibfield  {author} {\bibinfo {author} {\bibfnamefont {G.}~\bibnamefont
  {Galanti}}\ and\ \bibinfo {author} {\bibfnamefont {M.}~\bibnamefont
  {Roncadelli}},\ }\href {\doibase 10.3390/universe8050253} {\bibfield
  {journal} {\bibinfo  {journal} {Universe}\ }\textbf {\bibinfo {volume} {8}},\
  \bibinfo {pages} {253} (\bibinfo {year} {2022})},\ \Eprint
  {http://arxiv.org/abs/2205.00940} {arXiv:2205.00940 [hep-ph]} \BibitemShut
  {NoStop}%
\bibitem [{\citenamefont {Pshirkov}\ \emph {et~al.}(2011)\citenamefont
  {Pshirkov}, \citenamefont {Tinyakov}, \citenamefont {Kronberg},\ and\
  \citenamefont {Newton-McGee}}]{Pshirkov:2011um}%
  \BibitemOpen
  \bibfield  {author} {\bibinfo {author} {\bibfnamefont {M.~S.}\ \bibnamefont
  {Pshirkov}}, \bibinfo {author} {\bibfnamefont {P.~G.}\ \bibnamefont
  {Tinyakov}}, \bibinfo {author} {\bibfnamefont {P.~P.}\ \bibnamefont
  {Kronberg}}, \ and\ \bibinfo {author} {\bibfnamefont {K.~J.}\ \bibnamefont
  {Newton-McGee}},\ }\href {\doibase 10.1088/0004-637X/738/2/192} {\bibfield
  {journal} {\bibinfo  {journal} {Astrophys. J.}\ }\textbf {\bibinfo {volume}
  {738}},\ \bibinfo {pages} {192} (\bibinfo {year} {2011})},\ \Eprint
  {http://arxiv.org/abs/1103.0814} {arXiv:1103.0814 [astro-ph.GA]} \BibitemShut
  {NoStop}%
\bibitem [{\citenamefont {Jansson}\ and\ \citenamefont
  {Farrar}(2012)}]{Jansson:2012pc}%
  \BibitemOpen
  \bibfield  {author} {\bibinfo {author} {\bibfnamefont {R.}~\bibnamefont
  {Jansson}}\ and\ \bibinfo {author} {\bibfnamefont {G.~R.}\ \bibnamefont
  {Farrar}},\ }\href {\doibase 10.1088/0004-637X/757/1/14} {\bibfield
  {journal} {\bibinfo  {journal} {Astrophys. J.}\ }\textbf {\bibinfo {volume}
  {757}},\ \bibinfo {pages} {14} (\bibinfo {year} {2012})},\ \Eprint
  {http://arxiv.org/abs/1204.3662} {arXiv:1204.3662 [astro-ph.GA]} \BibitemShut
  {NoStop}%
\bibitem [{\citenamefont {{Taylor}}\ and\ \citenamefont
  {{Perley}}(1993)}]{1993ApJ416554T}%
  \BibitemOpen
  \bibfield  {author} {\bibinfo {author} {\bibfnamefont {G.~B.}\ \bibnamefont
  {{Taylor}}}\ and\ \bibinfo {author} {\bibfnamefont {R.~A.}\ \bibnamefont
  {{Perley}}},\ }\href {\doibase 10.1086/173257} {\bibfield  {journal}
  {\bibinfo  {journal} {\apj}\ }\textbf {\bibinfo {volume} {416}},\ \bibinfo
  {pages} {554} (\bibinfo {year} {1993})}\BibitemShut {NoStop}%
\bibitem [{\citenamefont {Ensslin}\ and\ \citenamefont
  {Vogt}(2003)}]{Ensslin:2003ez}%
  \BibitemOpen
  \bibfield  {author} {\bibinfo {author} {\bibfnamefont {T.~A.}\ \bibnamefont
  {Ensslin}}\ and\ \bibinfo {author} {\bibfnamefont {C.}~\bibnamefont {Vogt}},\
  }\href {\doibase 10.1051/0004-6361:20030172} {\bibfield  {journal} {\bibinfo
  {journal} {Astron. Astrophys.}\ }\textbf {\bibinfo {volume} {401}},\ \bibinfo
  {pages} {835} (\bibinfo {year} {2003})},\ \Eprint
  {http://arxiv.org/abs/astro-ph/0302426} {arXiv:astro-ph/0302426} \BibitemShut
  {NoStop}%
\bibitem [{\citenamefont {Vogt}\ and\ \citenamefont
  {Ensslin}(2003)}]{Vogt:2003su}%
  \BibitemOpen
  \bibfield  {author} {\bibinfo {author} {\bibfnamefont {C.}~\bibnamefont
  {Vogt}}\ and\ \bibinfo {author} {\bibfnamefont {T.~A.}\ \bibnamefont
  {Ensslin}},\ }\href {\doibase 10.1051/0004-6361:20031434} {\bibfield
  {journal} {\bibinfo  {journal} {Astron. Astrophys.}\ }\textbf {\bibinfo
  {volume} {412}},\ \bibinfo {pages} {373} (\bibinfo {year} {2003})},\ \Eprint
  {http://arxiv.org/abs/astro-ph/0309441} {arXiv:astro-ph/0309441} \BibitemShut
  {NoStop}%
\bibitem [{\citenamefont {Horns}\ \emph {et~al.}(2012)\citenamefont {Horns},
  \citenamefont {Maccione}, \citenamefont {Meyer}, \citenamefont {Mirizzi},
  \citenamefont {Montanino},\ and\ \citenamefont {Roncadelli}}]{Horns:2012kw}%
  \BibitemOpen
  \bibfield  {author} {\bibinfo {author} {\bibfnamefont {D.}~\bibnamefont
  {Horns}}, \bibinfo {author} {\bibfnamefont {L.}~\bibnamefont {Maccione}},
  \bibinfo {author} {\bibfnamefont {M.}~\bibnamefont {Meyer}}, \bibinfo
  {author} {\bibfnamefont {A.}~\bibnamefont {Mirizzi}}, \bibinfo {author}
  {\bibfnamefont {D.}~\bibnamefont {Montanino}}, \ and\ \bibinfo {author}
  {\bibfnamefont {M.}~\bibnamefont {Roncadelli}},\ }\href {\doibase
  10.1103/PhysRevD.86.075024} {\bibfield  {journal} {\bibinfo  {journal} {Phys.
  Rev. D}\ }\textbf {\bibinfo {volume} {86}},\ \bibinfo {pages} {075024}
  (\bibinfo {year} {2012})},\ \Eprint {http://arxiv.org/abs/1207.0776}
  {arXiv:1207.0776 [astro-ph.HE]} \BibitemShut {NoStop}%
\bibitem [{\citenamefont {Govoni}\ and\ \citenamefont
  {Feretti}(2004)}]{Govoni:2004as}%
  \BibitemOpen
  \bibfield  {author} {\bibinfo {author} {\bibfnamefont {F.}~\bibnamefont
  {Govoni}}\ and\ \bibinfo {author} {\bibfnamefont {L.}~\bibnamefont
  {Feretti}},\ }\href {\doibase 10.1142/S0218271804005080} {\bibfield
  {journal} {\bibinfo  {journal} {Int. J. Mod. Phys. D}\ }\textbf {\bibinfo
  {volume} {13}},\ \bibinfo {pages} {1549} (\bibinfo {year} {2004})},\ \Eprint
  {http://arxiv.org/abs/astro-ph/0410182} {arXiv:astro-ph/0410182} \BibitemShut
  {NoStop}%
\bibitem [{\citenamefont {Meyer}\ \emph {et~al.}(2013)\citenamefont {Meyer},
  \citenamefont {Horns},\ and\ \citenamefont {Raue}}]{Meyer:2013pny}%
  \BibitemOpen
  \bibfield  {author} {\bibinfo {author} {\bibfnamefont {M.}~\bibnamefont
  {Meyer}}, \bibinfo {author} {\bibfnamefont {D.}~\bibnamefont {Horns}}, \ and\
  \bibinfo {author} {\bibfnamefont {M.}~\bibnamefont {Raue}},\ }\href {\doibase
  10.1103/PhysRevD.87.035027} {\bibfield  {journal} {\bibinfo  {journal} {Phys.
  Rev. D}\ }\textbf {\bibinfo {volume} {87}},\ \bibinfo {pages} {035027}
  (\bibinfo {year} {2013})},\ \Eprint {http://arxiv.org/abs/1302.1208}
  {arXiv:1302.1208 [astro-ph.HE]} \BibitemShut {NoStop}%
\bibitem [{\citenamefont {Vikhlinin}\ \emph {et~al.}(2006)\citenamefont
  {Vikhlinin}, \citenamefont {Kravtsov}, \citenamefont {Forman}, \citenamefont
  {Jones}, \citenamefont {Markevitch}, \citenamefont {Murray},\ and\
  \citenamefont {Van~Speybroeck}}]{Vikhlinin:2005mp}%
  \BibitemOpen
  \bibfield  {author} {\bibinfo {author} {\bibfnamefont {A.}~\bibnamefont
  {Vikhlinin}}, \bibinfo {author} {\bibfnamefont {A.}~\bibnamefont {Kravtsov}},
  \bibinfo {author} {\bibfnamefont {W.}~\bibnamefont {Forman}}, \bibinfo
  {author} {\bibfnamefont {C.}~\bibnamefont {Jones}}, \bibinfo {author}
  {\bibfnamefont {M.}~\bibnamefont {Markevitch}}, \bibinfo {author}
  {\bibfnamefont {S.~S.}\ \bibnamefont {Murray}}, \ and\ \bibinfo {author}
  {\bibfnamefont {L.}~\bibnamefont {Van~Speybroeck}},\ }\href {\doibase
  10.1086/500288} {\bibfield  {journal} {\bibinfo  {journal} {Astrophys. J.}\
  }\textbf {\bibinfo {volume} {640}},\ \bibinfo {pages} {691} (\bibinfo {year}
  {2006})},\ \Eprint {http://arxiv.org/abs/astro-ph/0507092}
  {arXiv:astro-ph/0507092} \BibitemShut {NoStop}%
\bibitem [{\citenamefont {Abuter}\ \emph {et~al.}(2019)\citenamefont {Abuter}
  \emph {et~al.}}]{Gravity:2019nxk}%
  \BibitemOpen
  \bibfield  {author} {\bibinfo {author} {\bibfnamefont {R.}~\bibnamefont
  {Abuter}} \emph {et~al.} (\bibinfo {collaboration} {Gravity}),\ }\href
  {\doibase 10.1051/0004-6361/201935656} {\bibfield  {journal} {\bibinfo
  {journal} {Astron. Astrophys.}\ }\textbf {\bibinfo {volume} {625}} (\bibinfo
  {year} {2019}),\ 10.1051/0004-6361/201935656},\ \Eprint
  {http://arxiv.org/abs/1904.05721} {arXiv:1904.05721 [astro-ph.GA]}
  \BibitemShut {NoStop}%
\bibitem [{\citenamefont {Beck}(2009)}]{Beck:2008ty}%
  \BibitemOpen
  \bibfield  {author} {\bibinfo {author} {\bibfnamefont {R.}~\bibnamefont
  {Beck}},\ }\href {\doibase 10.1063/1.3076806} {\bibfield  {journal} {\bibinfo
   {journal} {AIP Conf. Proc.}\ }\textbf {\bibinfo {volume} {1085}},\ \bibinfo
  {pages} {83} (\bibinfo {year} {2009})},\ \Eprint
  {http://arxiv.org/abs/0810.2923} {arXiv:0810.2923 [astro-ph]} \BibitemShut
  {NoStop}%
\bibitem [{\citenamefont {Cordes}\ and\ \citenamefont
  {Lazio}(2002)}]{Cordes:2002wz}%
  \BibitemOpen
  \bibfield  {author} {\bibinfo {author} {\bibfnamefont {J.~M.}\ \bibnamefont
  {Cordes}}\ and\ \bibinfo {author} {\bibfnamefont {T.~J.~W.}\ \bibnamefont
  {Lazio}},\ }\href@noop {} {\  (\bibinfo {year} {2002})},\ \Eprint
  {http://arxiv.org/abs/astro-ph/0207156} {arXiv:astro-ph/0207156} \BibitemShut
  {NoStop}%
\bibitem [{\citenamefont {Ade}\ \emph {et~al.}(2016)\citenamefont {Ade} \emph
  {et~al.}}]{Planck:2015zrl}%
  \BibitemOpen
  \bibfield  {author} {\bibinfo {author} {\bibfnamefont {P.~A.~R.}\
  \bibnamefont {Ade}} \emph {et~al.} (\bibinfo {collaboration} {Planck}),\
  }\href {\doibase 10.1051/0004-6361/201525821} {\bibfield  {journal} {\bibinfo
   {journal} {Astron. Astrophys.}\ }\textbf {\bibinfo {volume} {594}},\
  \bibinfo {pages} {A19} (\bibinfo {year} {2016})},\ \Eprint
  {http://arxiv.org/abs/1502.01594} {arXiv:1502.01594 [astro-ph.CO]}
  \BibitemShut {NoStop}%
\bibitem [{\citenamefont {Pshirkov}\ \emph {et~al.}(2016)\citenamefont
  {Pshirkov}, \citenamefont {Tinyakov},\ and\ \citenamefont
  {Urban}}]{Pshirkov:2015tua}%
  \BibitemOpen
  \bibfield  {author} {\bibinfo {author} {\bibfnamefont {M.~S.}\ \bibnamefont
  {Pshirkov}}, \bibinfo {author} {\bibfnamefont {P.~G.}\ \bibnamefont
  {Tinyakov}}, \ and\ \bibinfo {author} {\bibfnamefont {F.~R.}\ \bibnamefont
  {Urban}},\ }\href {\doibase 10.1103/PhysRevLett.116.191302} {\bibfield
  {journal} {\bibinfo  {journal} {Phys. Rev. Lett.}\ }\textbf {\bibinfo
  {volume} {116}},\ \bibinfo {pages} {191302} (\bibinfo {year} {2016})},\
  \Eprint {http://arxiv.org/abs/1504.06546} {arXiv:1504.06546 [astro-ph.CO]}
  \BibitemShut {NoStop}%
\bibitem [{\citenamefont {Ackermann}\ \emph {et~al.}(2015)\citenamefont
  {Ackermann} \emph {et~al.}}]{Fermi-LAT:2014ryh}%
  \BibitemOpen
  \bibfield  {author} {\bibinfo {author} {\bibfnamefont {M.}~\bibnamefont
  {Ackermann}} \emph {et~al.} (\bibinfo {collaboration} {Fermi-LAT}),\ }\href
  {\doibase 10.1088/0004-637X/799/1/86} {\bibfield  {journal} {\bibinfo
  {journal} {Astrophys. J.}\ }\textbf {\bibinfo {volume} {799}},\ \bibinfo
  {pages} {86} (\bibinfo {year} {2015})},\ \Eprint
  {http://arxiv.org/abs/1410.3696} {arXiv:1410.3696 [astro-ph.HE]} \BibitemShut
  {NoStop}%
\bibitem [{\citenamefont {Weidenspointner}\ \emph {et~al.}(2000)\citenamefont
  {Weidenspointner} \emph {et~al.}}]{Weidenspointner:2000aq}%
  \BibitemOpen
  \bibfield  {author} {\bibinfo {author} {\bibfnamefont {G.}~\bibnamefont
  {Weidenspointner}} \emph {et~al.},\ }\href {\doibase 10.1063/1.1303269}
  {\bibfield  {journal} {\bibinfo  {journal} {AIP Conf. Proc.}\ }\textbf
  {\bibinfo {volume} {510}},\ \bibinfo {pages} {581} (\bibinfo {year}
  {2000})},\ \Eprint {http://arxiv.org/abs/astro-ph/0012332}
  {arXiv:astro-ph/0012332} \BibitemShut {NoStop}%
\bibitem [{\citenamefont {Tsygankov}\ \emph {et~al.}(2007)\citenamefont
  {Tsygankov}, \citenamefont {Lutovinov}, \citenamefont {Churazov},\ and\
  \citenamefont {Sunyaev}}]{Tsygankov:2007jf}%
  \BibitemOpen
  \bibfield  {author} {\bibinfo {author} {\bibfnamefont {S.~S.}\ \bibnamefont
  {Tsygankov}}, \bibinfo {author} {\bibfnamefont {A.~A.}\ \bibnamefont
  {Lutovinov}}, \bibinfo {author} {\bibfnamefont {E.~M.}\ \bibnamefont
  {Churazov}}, \ and\ \bibinfo {author} {\bibfnamefont {R.~A.}\ \bibnamefont
  {Sunyaev}},\ }\href {\doibase 10.1134/S1063773707060023} {\bibfield
  {journal} {\bibinfo  {journal} {Astron. Lett.}\ }\textbf {\bibinfo {volume}
  {33}},\ \bibinfo {pages} {368} (\bibinfo {year} {2007})},\ \Eprint
  {http://arxiv.org/abs/0704.2874} {arXiv:0704.2874 [astro-ph]} \BibitemShut
  {NoStop}%
\bibitem [{\citenamefont {Bartels}\ \emph {et~al.}(2017)\citenamefont
  {Bartels}, \citenamefont {Gaggero},\ and\ \citenamefont
  {Weniger}}]{Bartels:2017dpb}%
  \BibitemOpen
  \bibfield  {author} {\bibinfo {author} {\bibfnamefont {R.}~\bibnamefont
  {Bartels}}, \bibinfo {author} {\bibfnamefont {D.}~\bibnamefont {Gaggero}}, \
  and\ \bibinfo {author} {\bibfnamefont {C.}~\bibnamefont {Weniger}},\ }\href
  {\doibase 10.1088/1475-7516/2017/05/001} {\bibfield  {journal} {\bibinfo
  {journal} {JCAP}\ }\textbf {\bibinfo {volume} {05}},\ \bibinfo {pages} {001}
  (\bibinfo {year} {2017})},\ \Eprint {http://arxiv.org/abs/1703.02546}
  {arXiv:1703.02546 [astro-ph.HE]} \BibitemShut {NoStop}%
\bibitem [{\citenamefont {Ballesteros}\ \emph {et~al.}(2020)\citenamefont
  {Ballesteros}, \citenamefont {Coronado-Bl{\'a}zquez},\ and\ \citenamefont
  {Gaggero}}]{Ballesteros:2019exr}%
  \BibitemOpen
  \bibfield  {author} {\bibinfo {author} {\bibfnamefont {G.}~\bibnamefont
  {Ballesteros}}, \bibinfo {author} {\bibfnamefont {J.}~\bibnamefont
  {Coronado-Bl{\'a}zquez}}, \ and\ \bibinfo {author} {\bibfnamefont
  {D.}~\bibnamefont {Gaggero}},\ }\href {\doibase
  10.1016/j.physletb.2020.135624} {\bibfield  {journal} {\bibinfo  {journal}
  {Phys. Lett. B}\ }\textbf {\bibinfo {volume} {808}},\ \bibinfo {pages}
  {135624} (\bibinfo {year} {2020})},\ \Eprint
  {http://arxiv.org/abs/1906.10113} {arXiv:1906.10113 [astro-ph.CO]}
  \BibitemShut {NoStop}%
\bibitem [{\citenamefont {Edwards}\ and\ \citenamefont
  {Weniger}(2018)}]{Edwards:2017mnf}%
  \BibitemOpen
  \bibfield  {author} {\bibinfo {author} {\bibfnamefont {T.~D.~P.}\
  \bibnamefont {Edwards}}\ and\ \bibinfo {author} {\bibfnamefont
  {C.}~\bibnamefont {Weniger}},\ }\href {\doibase
  10.1088/1475-7516/2018/02/021} {\bibfield  {journal} {\bibinfo  {journal}
  {JCAP}\ }\textbf {\bibinfo {volume} {02}},\ \bibinfo {pages} {021} (\bibinfo
  {year} {2018})},\ \Eprint {http://arxiv.org/abs/1704.05458} {arXiv:1704.05458
  [astro-ph.IM]} \BibitemShut {NoStop}%
\bibitem [{\citenamefont {Anastassopoulos}\ \emph {et~al.}(2017)\citenamefont
  {Anastassopoulos} \emph {et~al.}}]{CAST:2017uph}%
  \BibitemOpen
  \bibfield  {author} {\bibinfo {author} {\bibfnamefont {V.}~\bibnamefont
  {Anastassopoulos}} \emph {et~al.} (\bibinfo {collaboration} {CAST}),\ }\href
  {\doibase 10.1038/nphys4109} {\bibfield  {journal} {\bibinfo  {journal}
  {Nature Phys.}\ }\textbf {\bibinfo {volume} {13}},\ \bibinfo {pages} {584}
  (\bibinfo {year} {2017})},\ \Eprint {http://arxiv.org/abs/1705.02290}
  {arXiv:1705.02290 [hep-ex]} \BibitemShut {NoStop}%
\bibitem [{\citenamefont {O'Hare}(2020)}]{AxionLimits}%
  \BibitemOpen
  \bibfield  {author} {\bibinfo {author} {\bibfnamefont {C.}~\bibnamefont
  {O'Hare}},\ }\href {\doibase 10.5281/zenodo.3932430} {\enquote {\bibinfo
  {title} {cajohare/axionlimits: Axionlimits},}\ }\bibinfo {howpublished}
  {\url{https://cajohare.github.io/AxionLimits/}} (\bibinfo {year}
  {2020})\BibitemShut {NoStop}%
\end{thebibliography}%

\end{document}